\documentclass[preprint,3p]{elsarticle}

\usepackage{amssymb}
\usepackage{times}
\usepackage{graphicx}
\usepackage{dcolumn}
\usepackage{bm}
\usepackage{epsfig}
\usepackage{epstopdf}
\usepackage{natbib}
\usepackage[version=4]{mhchem}
\usepackage{soul}
\usepackage[mathlines]{lineno}
\usepackage{color,soul}
\usepackage{xfrac}
\usepackage[font=normalsize,labelfont=bf]{caption}

\newcommand{\bra}[1]{\left(#1\right)}

\journal{Physica D}

\begin{document}
	
	\begin{frontmatter}
		
		
		
		
		\title{Spatial asymmetries of resonant oscillations in periodically forced heterogeneous media}

		\author[mymainaddress]{Yuval Edri\corref{mycurrentaddress}}
		\author[mymainaddress,aysecondaryaddress]{Ehud Meron}
		\author[mymainaddress,aysecondaryaddress]{Arik Yochelis\corref{mycorrespondingauthor}}
		
		\cortext[mycorrespondingauthor]{Corresponding author}
		\cortext[mycurrentaddress]{Current address: Laboratory of Sensory Neuroscience, The Rockefeller University, New York, NY 10065, USA}
		
		\address[mymainaddress]{Department of Solar Energy and Environmental Physics, Swiss Institute for Dryland Environmental and Energy Research, Blaustein Institutes for Desert Research, Ben-Gurion University of the Negev, Sede Boqer Campus, 8499000 Midreshet Ben-Gurion, Israel}
		
		\address[aysecondaryaddress]{Department of Physics, Ben-Gurion University of the Negev, 8410501 Beer Sheva, Israel}
		
		\begin{abstract}
		{Spatially localized oscillations in periodically forced systems are intriguing phenomena. They may occur in spatially homogeneous media (oscillons), but quite often emerge in heterogeneous media, such as the auditory system, where {localized} oscillations are believed to play an important role in frequency discrimination of incoming sound waves. In this paper, we use an amplitude-equation approach to study the {spatial profile of the oscillations} and the factors that affect it. More specifically, we use a variant of the forced complex Ginzburg-Landau (FCGL) equation {to} describes an oscillatory system below the Hopf bifurcation with space-dependent Hopf frequency, subject to both parametric and additive forcing. We show that spatial heterogeneity, combined with bistability of system states, results in {spatial }asymmetry of the localized oscillations. We further identify parameters that control that asymmetry, and characterize the spatial profile of the oscillations in terms of {maximum} amplitude, location, width and asymmetry. Our results {bear} qualitative similarities to empirical observation trends that have found in the auditory system.}
		\end{abstract}
		
		
	\end{frontmatter}
	
	\vskip 0.5in
    Highlights
    \begin{itemize}
    	\item The study is motivated by localized oscillations observed in the cochlea \\
    	\item Heterogenous oscillatory media under parametric and additive forcing are considered \\
    	\item Spatial profiles of localized oscillations are studied analytically and numerically \\
    	\item Factors controlling the asymmetry of the oscillations profile are identified \\
    	\item Relations to observations in the cochlea are discussed
    \end{itemize}
	\newpage
	\section{Introduction}
	
	Spatially localized resonant oscillations, such as oscillons~\cite{longhi1996stable,umbanhowar1996localized,pradenas2017slanted} and reciprocal oscillons~\cite{petrov1997resonant,blair2000patterns} are not only inspiring natural phenomena but also mathematically intriguing~\cite{Elphick1989pra,YochelisBurkeKnobloch2006,BurkeYochelisKnobloch2008,goddard2009fluid,kenig2009intrinsic,dawes2010localized,MaBurkeKnobloch2010,noskov2012oscillons,mcquighan2014oscillons,alnahdi2014localized,ma2016two,castiloyochelis}. Similarly to most frequency-locking studies of oscillatory media~\cite{petrov1997resonant,lin2000resonant,melo1995hexagons,melo1994transition,arbell2000temporally,shats2012parametrically,coullet1994excitable,tlidi1998kinetics,izus2000bloch,imbihl1995oscillatory,Lin2004pre,Marts2004prl}, also localized resonances have been studied {under the assumption of} spatially homogeneous conditions. However, in some cases spatial heterogeneity is an inherent feature of the system and is paramount to the emergence of distinct type of localized oscillations. Examples are the cochlea of the inner ear~\cite{pickles_book,dallos1996overview,kandel}, alligator water dance~\cite{moriarty2011faraday}, and Faraday waves under heterogeneous parametric excitation~\cite{urra2019localized}. In other cases, heterogeneities can be exploited to tune the performance of particular engineered outputs in a range of potential applications, including mechanical resonators~\cite{lifshitz2010nonlinear,jia2013parametrically,abrams2014nonlinear}, catalytic surface reactions~\cite{imbihl1995oscillatory}, and plasmonic nanoparticles~\cite{noskov2012oscillons}.
		
	{This work is motivated by the spatial heterogeneity of the cochlea and its resonant response to incoming sound waves. It was Helmholtz in 1863, who suggested a heuristic resonant `inverse piano' realization to sound discrimination in cochlea~\cite{helmholtz2013sensations}. The underlying approach was the consideration of uncoupled oscillators with frequencies that vary in a monotonic fashion in space, which resonates with an externally applied uniform periodic forcing that represented the incoming sound wave. However, an apparent flaw of the resulting resonant behavior stemmed from its inability to describe a traveling wave that, according to von B\'{e}k\'{e}sy~\cite{von_bekesy}, arises at the base of the cochlea and propagates up to a specific location along it, where that location depends on the sound-wave frequency. As a result, an alternative `traveling--wave' description emerged, suggesting that the localized response to incoming sound waves can be treated via a heterogeneous wave equation~\cite{lighthill1981energy,olson2012bekesy}. It was only after the discoveries of spontaneous (otoacoustic) emissions of sound waves from the ear~\cite{kemp} and of the nonlinear amplified responses to sound~\cite{ruggero97}, that the resonance approach has been brought back to spotlight~\cite{essential,vilfan2008frequency,hudspeth2010,fruth2014active,Hudspeth2014}. Moreover, it was Bell~\cite{bell2012resonance} who conjectured that the resonance approach is a plausible description also to the emerging localized traveling waves since these can be interpreted as phase waves, alternatively to the energy propagation view. It should be emphasized that localized oscillations in the cochlea~\cite{ourPRE} are distinct from oscillons in homogeneous media, where the localization is generally related to homoclinic snaking (regular, collapsed, or slanted)~\cite{knoblochreview} and the references therein. In the cochlea, the localization is a direct consequence of a spatially dependent resonant response~\cite{ourChaos} and is not associated with homoclinic connections inside the resonance region~\cite{BurkeYochelisKnobloch2008,MaBurkeKnobloch2010}. Moreover, profiles of localized vibration that were measured at different locations along the cochlea present different asymmetries in their shape~\cite{RoblesRuggero2001,reichenbach2010ratchet,olson2012bekesy}.
	
	In two companion papers, we {have} studied the effect of spatial heterogeneity on the asymmetry of resonant responses in the presence of combined additive and parametric forcing~\cite{ourPRE} and on the shape of the resonance boundary~\cite{ourChaos}. The studies exploited the forced complex Ginzburg-Landau (FCGL) equation that is universal near the Hopf instability~\cite{essential}. It was found that when the forcing is introduced to the entire domain, the envelope forms, in general, do not depend on whether the forcing is uniform or of a traveling wave form~\cite{ourPRE}. The FCGL framework aligns with the resonance approach for the cochlea~\cite{bell2012resonance,ammari2019fully}, including the capability to address other resonance orders~\cite{levy2016high,ammari2019fully}. Yet, FCGL should not be regarded as a model for the cochlea but as a means to gain insights into specific aspects of localized resonant oscillations.}	
		
	In this study, we extend the results of earlier works~\cite{ourPRE,ourChaos}, addressing the effects of nonlinearity and bistability on the spatial form of localized resonant oscillations. We continue to apply the FCGL equation to the 1:1 resonance, where the system oscillates at exactly the forcing frequency (Section~\ref{sec:FCGL}), taking into consideration a monotonic dependence of the unforced frequency on the space coordinate. The analysis focuses on the impact of the nonlinear frequency correction term, which is essential for bistability of resonant solutions of low amplitude (possibly zero) and high amplitude. We show that in the absence of bistability, the asymmetry of the spatial profile of localized oscillations can be derived by studying the spatially uncoupled system (Sections~\ref{sec:uncoupled} and~\ref{sec:profile_nu}) while in the presence of bistability, the asymmetry is determined by a crossover point between the coexisting solutions, {which is} related to the Maxwell point of front solutions of the homogeneous system (Section~\ref{sec:spatial}). We further show that the results apply to pure parametric, pure additive, or combined driving force, and that it can be generalized to variations of other parameters, such as the distance from the Hopf onset.
	
	\section{{Amplitude} equation approach {to} forced oscillations and localized profiles}\label{sec:FCGL}
	
	Resonant behavior in a spatially extended oscillatory medium driven by additive and parametric forcing is well described by the FCGL amplitude equation~\cite{CoulletEmilsson1992,meron2015nonlinear}, where for 1:1 resonance it reads~\cite{lifshitz,ourepl}:
	\begin{equation}\label{eq:FCGL}
	\frac{\partial A}{\partial t}=\left(\mu+i\nu \right)A-(1+i\beta)|A|^2A+\Gamma_p \bar{A}+\Gamma_a+(1+i\alpha)\nabla^2A.
	\end{equation}
	Here, $\bar{A}$ is the complex conjugate of $A$, and the parameters $\mu,\nu,\beta,\alpha \in \mathbb{R}$ describe, respectively, the distance from the Hopf bifurcation, deviation of the forcing frequency from the unforced frequency (hereafter ``detuning''), nonlinear frequency correction and dispersion, and $\Gamma_p\in \mathbb{C}$, $\Gamma_a\in \mathbb{R}$ are related to the parametric and additive forcing, respectively.
	
	In an earlier paper~\cite{ourPRE}, we have shown that~\eqref{eq:FCGL} monotonic spatial inhomogeneity gives rise, in one space dimension (1D), to distinct asymmetric shapes of localized resonances, emphasizing the relative impact of parametric forcing, $\Gamma_p$. In what follows, we wish to complete this analysis by focusing on conditions that give rise to bistability of resonant solutions (obtained with $\beta\ne 0$), confining ourselves to damped oscillations, that is, $\mu<0$ (specifically we use throughout all the computations $\mu=-0.05$). For consistency with Ref.~\cite{ourChaos}, we use the following simplified version of~\eqref{eq:FCGL},
	\begin{equation}\label{eq:FCGLmodel}
	\frac{\partial A}{\partial t}=\bra{\mu+i\nu(x)}A-\left(1+i\beta\right)|A|^2A+\Gamma_a+\Gamma_{p}\bar A
	+D\frac{\partial^2 A}{\partial x^2}\,,
	\end{equation}
	where we assume a \emph{linear} spatial dependence of the detuning, $\nu(x)=2\eta x+\nu_0$ where $\nu_0\in \mathbb{R}$ is the detuning of the homogeneous system, rescale the spatial range to unity, $x\in[-\sfrac{1}{2},\sfrac{1}{2}]$, and choose {$\alpha=0$} and $0<\eta,\Gamma_p,D\in \mathbb{R}$. Note that we also consider spatially uniform forcing and not a traveling wave as in~\cite{ourPRE}, {where an advective term is added to~\eqref{eq:FCGLmodel}}.	
	
\begin{figure}[tp]
		\centering
		(a)\includegraphics[width=0.4\linewidth]{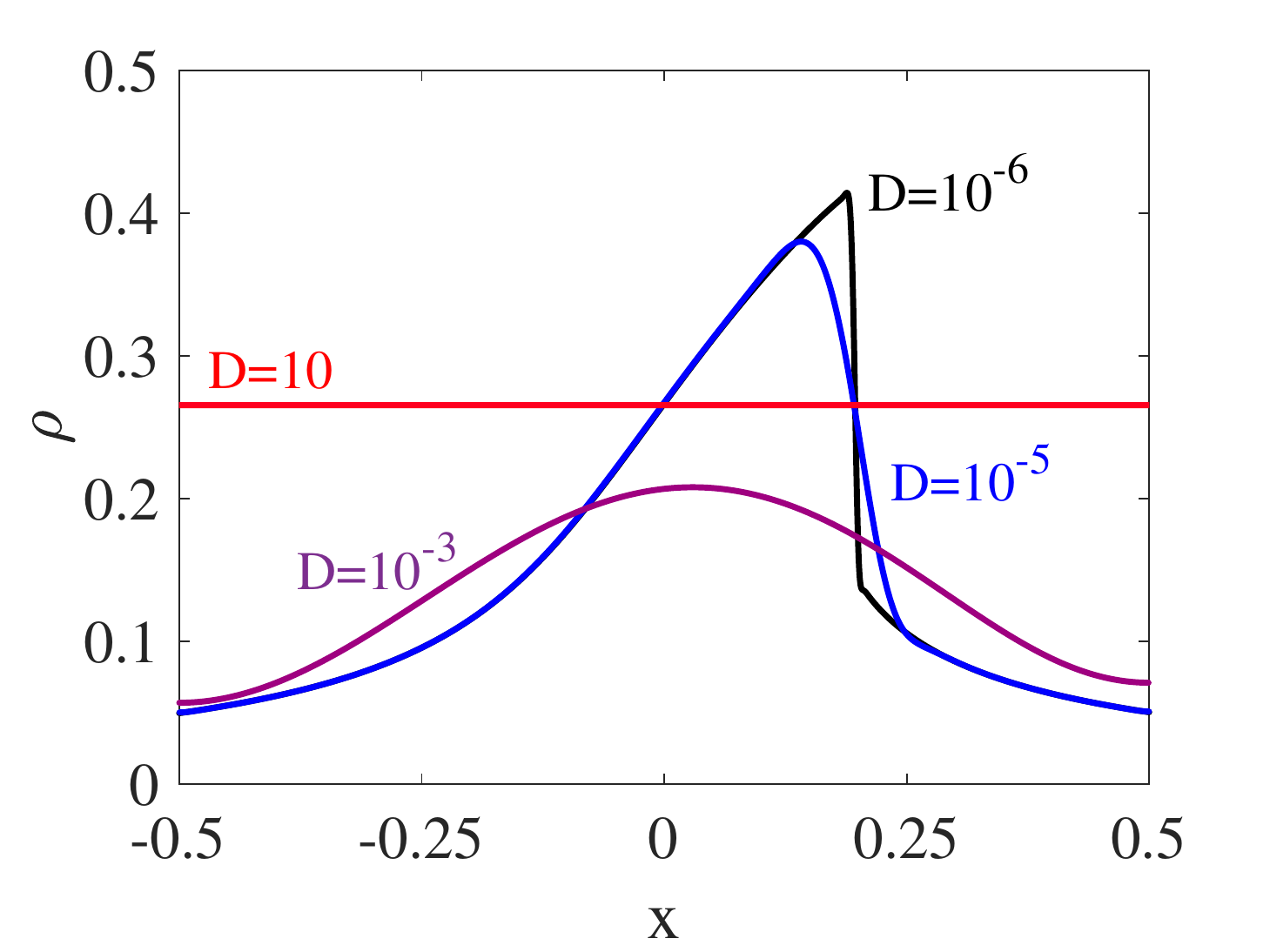}
		(b)\includegraphics[width=0.4\linewidth]{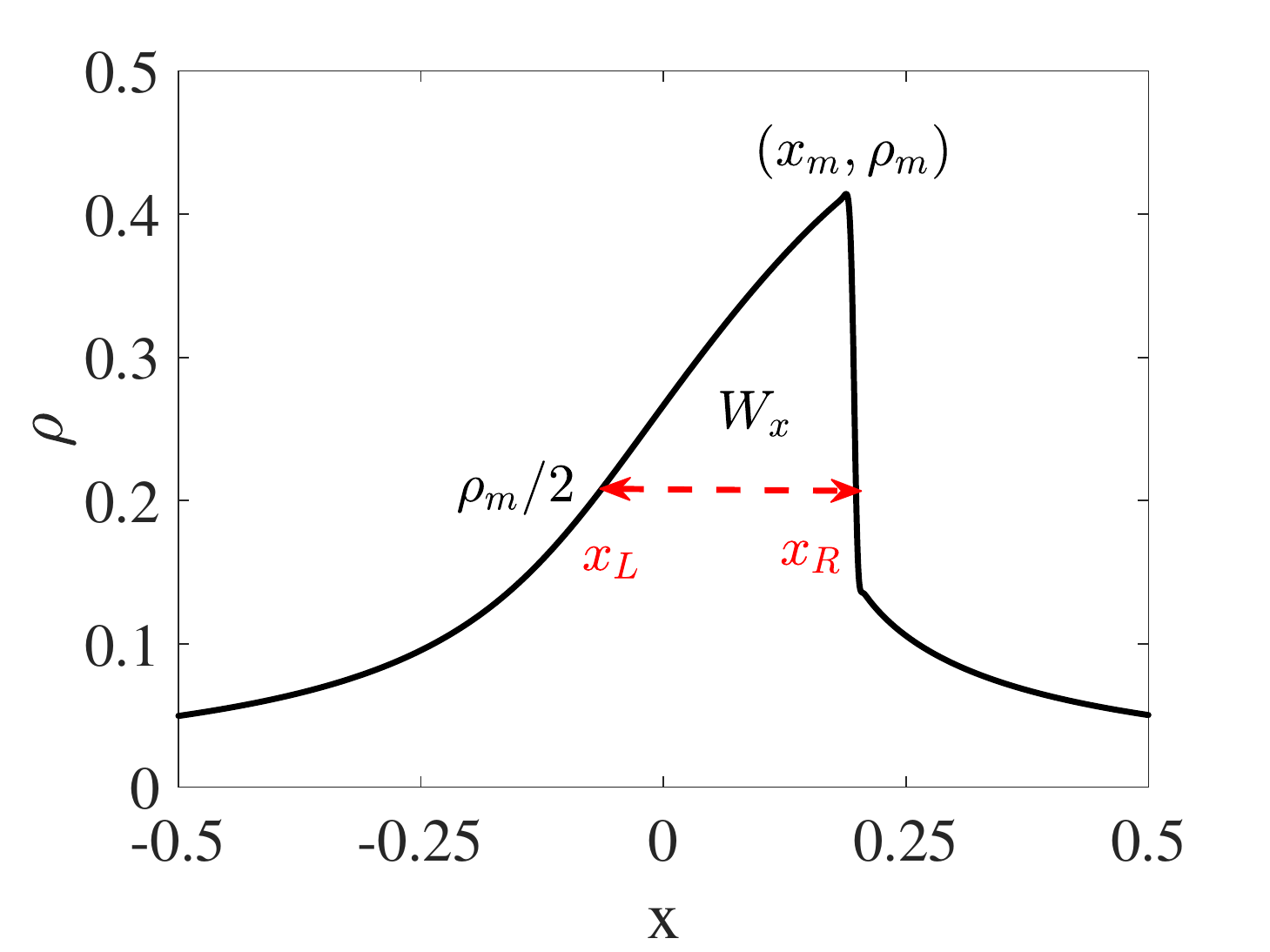}
		\caption{(a) Typical solution profiles of~\eqref{eq:FCGLmodel} for pure additive forcing and several $D$ values, in terms of the amplitude $\rho(x)=|A|$. The results were obtained numerically using the continuation package AUTO~\cite{doedel1998auto}. (b) Illustration of the profile attributes used in the definition of the asymmetry measure $\Lambda$ in ~\eqref{eq:assymetryMeasure}: $\rho_m$ maximum amplitude of the profile, $x_m$ location in space at which $\rho=\rho_m$, $W_x$ is the profile width at which $\rho=\rho_m/2$, and $x_L$ and $x_R$ are, respectively, left and right limits of the profile width calculated at $\rho=\rho_m/2$. Parameters: $\mu=-0.05$, $\nu_0=\Gamma_p=0$, $\beta=5$, $\Gamma_a=0.1$, $\eta=1/2$.}
		\label{fig:RhoVsD}
	\end{figure}

	The parameter $D$ quantifies the strength of the spatial coupling between nearby oscillatory elements and effectively affects the synchronization level~\cite{ourChaos}. For $D\gg1$ the whole spatial domain is synchronized so that the oscillation amplitude $\rho\equiv|A|$, where $A=\rho \exp{(i\phi)}$, is uniform throughout the domain, as shown in Fig.~\ref{fig:RhoVsD}. This result is demonstrated for pure additive forcing, {but similar results are obtained for forcing that contains a parametric component too}~\cite{ourChaos}. When $D\ll1$ spatial localization emerges, as Figure~\ref{fig:RhoVsD} shows. Note that the asymmetry of the localized profiles becomes prominent as $D$ is sufficiently decreased, turning into a sharp drop on the high $x$ side ($x\approx 0.2$) for $D=10^{-6}$.
	
	To characterize the spatial symmetry (or asymmetry) of the localized resonant profile, we find it useful to define the following measure:
	\begin{equation}\label{eq:assymetryMeasure}
	\Lambda=\int_{x_m}^{x_R}\rho^2(x)dx\Big/\int_{x_L}^{x_m}\rho^2(x)dx.
	\end{equation}
	where $x_m$ is the location in space at which the amplitude $\rho=|A|$ is maximal, i.e., $\rho=\rho_m$, and $x_L$ and $x_R$ are respective left and right limits of the peak width $W_x$ calculated at $\rho=\rho_m/2$, as shown in Fig.~\ref{fig:RhoVsD}(b). Notably, the peak location within $W_x$ defines the asymmetry, where for $\beta=0$ the localization has a symmetric shape, see also~\cite{ourChaos}. {We note that the measure~\eqref{eq:assymetryMeasure} is defined here in a different way compared to the similar measure defined in~\cite{ourPRE}, but still assumes the same limits: $\Lambda\to 1$ for symmetric profiles and $\Lambda\to 0$ for highly asymmetric profiles}. This new definition is more amenable to analytical explorations.
	
	In what follows, we study steady state solutions of~\eqref{eq:FCGLmodel} for parametric and additive forcing, and their dependence on the forcing amplitude and frequency (through the detuning parameter). These solutions represent 1:1 resonant oscillations. We find it useful to start with the uncoupled case, obtained by setting $D=0$, and relate the dependence of the solutions on the detuning parameter $\nu$ for $D=0$, to their spatial dependence for $D\ll 1$, by mapping $\nu \to x$. 
	
	\section{Coexistence of {stable} spatially-decoupled solutions for parametric and additive forcing}\label{sec:uncoupled}
	\subsection{Bistability regions}\label{sec:bistability}
	The amplitude $\rho=|A|$ of fixed points of~\eqref{eq:FCGLmodel} in spatially uncoupled systems ($D=0$), subjected to both parametric and additive forcing, solves the polynomial equation~\cite{ourepl}:
	\begin{equation}\label{eq:stat_amplitude}
	f(\nu)=(\rho^2-\mu)^2+\bra{\beta\rho^2-\nu}^2-\bra{1-\delta}\rho^{-2}\Gamma_a^2-\delta\cdot \Gamma_p^2=0,
	\end{equation}
	where
	\[
	\delta=\Bigg\{\begin{array}{cc}
	0, & \quad\text{for}\quad \Gamma_a >0,~~\Gamma_p=0\\
	1, & \quad\text{for}\quad \Gamma_a =0,~~\Gamma_p>0
	\end{array}.
	\]
	and the phase $\phi=\arg(A)$ solves the equation
	\[
	\cos{\bra{1+\delta}\phi}=\rho\frac{\rho^2-\mu}{\bra{1-\delta}|\Gamma_a|+\delta|\Gamma_p|\rho}\quad \text{or}\quad \sin{\bra{1+\delta}\phi}=\rho\frac{\nu-\beta\rho^2}{\bra{1-\delta}|\Gamma_a|+\delta|\Gamma_p|\rho}\,.
	\]
	
	The solutions of~\eqref{eq:stat_amplitude} for pure parametric forcing are~\cite{BurkeYochelisKnobloch2008}
	\begin{subequations}\label{eq:paraSOL}
		\begin{eqnarray}
		\rho^0_p&=&0,\\
		\rho_p^{\pm}&=&\sqrt{\frac{\mu+\beta\nu\pm\sqrt{\Gamma_p^2(1+\beta^2)-(\nu-\beta\mu)^2}}{1+\beta^2}},
		\end{eqnarray}
	\end{subequations}
	while for pure additive forcing they are:
	\begin{subequations}\label{eq:addSolutions}
		\begin{eqnarray}
		\label{eq:addSolutions_a}
		\rho_a^0&=&\frac{1}{\sqrt{3}}\sqrt{a+S_a^{1/3}-\left(3b-a^2\right)S_a^{-1/3}},\\
		\label{eq:addSolutions_b}
		\rho_a^{\pm}&=&\frac{1}{\sqrt{3}}\sqrt{a+\left(3b-a^2\right)\frac{1\pm i\sqrt{3}}{2}S_a^{-1/3}-\frac{1\mp i\sqrt{3}}{2}S_a^{1/3}},
		\end{eqnarray}
	\end{subequations}
	where
	\[
	S_a=\frac{2a^3-9ab+27c+3\sqrt{3}\sqrt{b^2(4b-a^2)+4a^3c-18abc+27c^2}}{2},~a=2\frac{\mu+\beta\nu}{1+\beta^2},~ b=\frac{\mu^2+\nu^2}{1+\beta^2},~ c=\frac{\Gamma_a^2}{1+\beta^2}
	\]
	
	Notably, for both parametric and additive forcing, Eq.~\eqref{eq:stat_amplitude} preserves the inversion symmetry $(\beta,\nu)\rightarrow(-\beta,-\nu)$~\cite{yipingtheis}. {In the following we explore bistability regions of fixed points in the parameter plane spanned by the detuning $\nu$ and the forcing amplitude, $\Gamma_p$ or $\Gamma_a$, for different values of $\beta>0$.}
	\begin{figure}[tp]
		\centering
		(a)\includegraphics[width=0.4\linewidth]{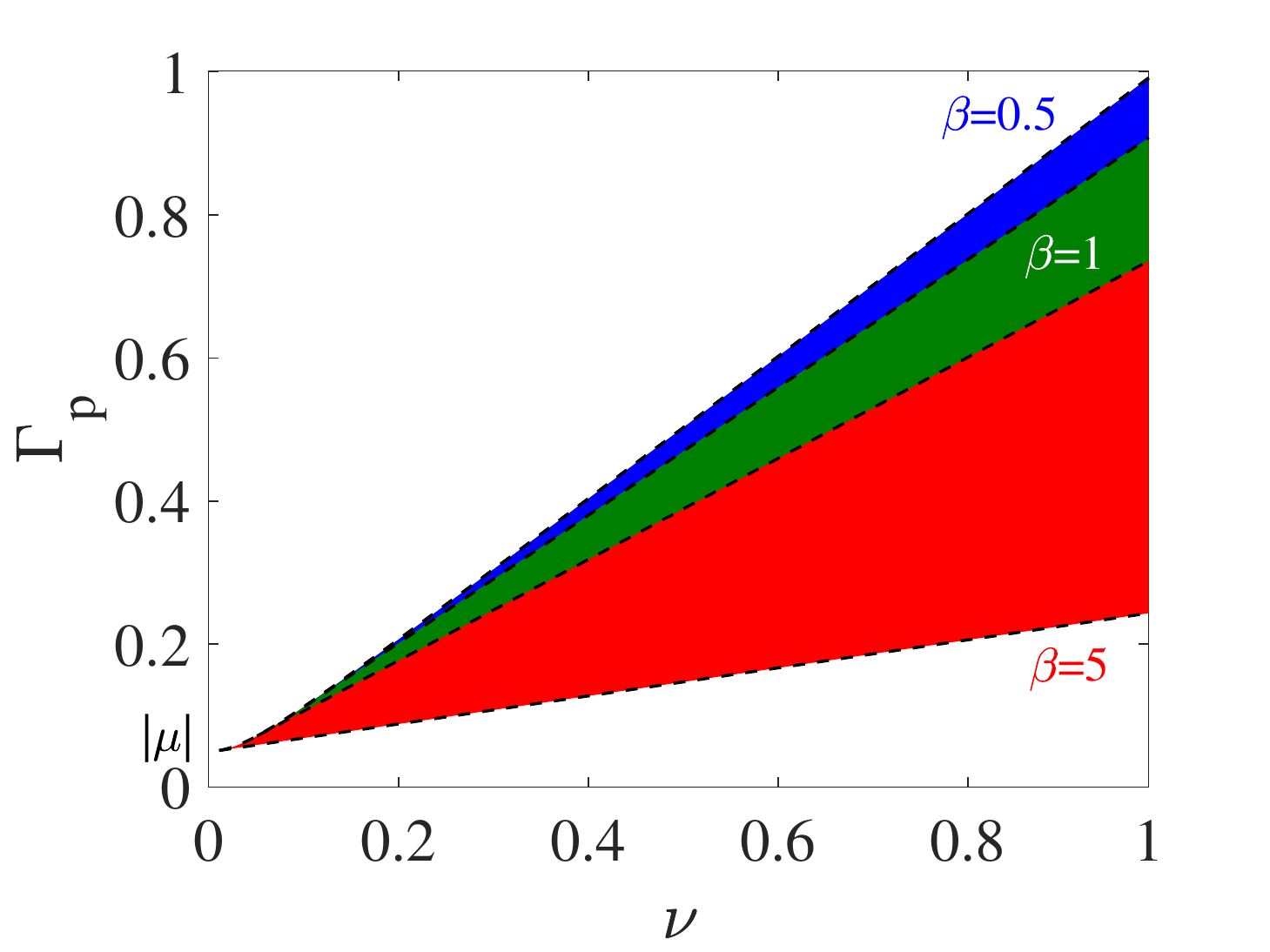}
		(b)\includegraphics[width=0.4\linewidth]{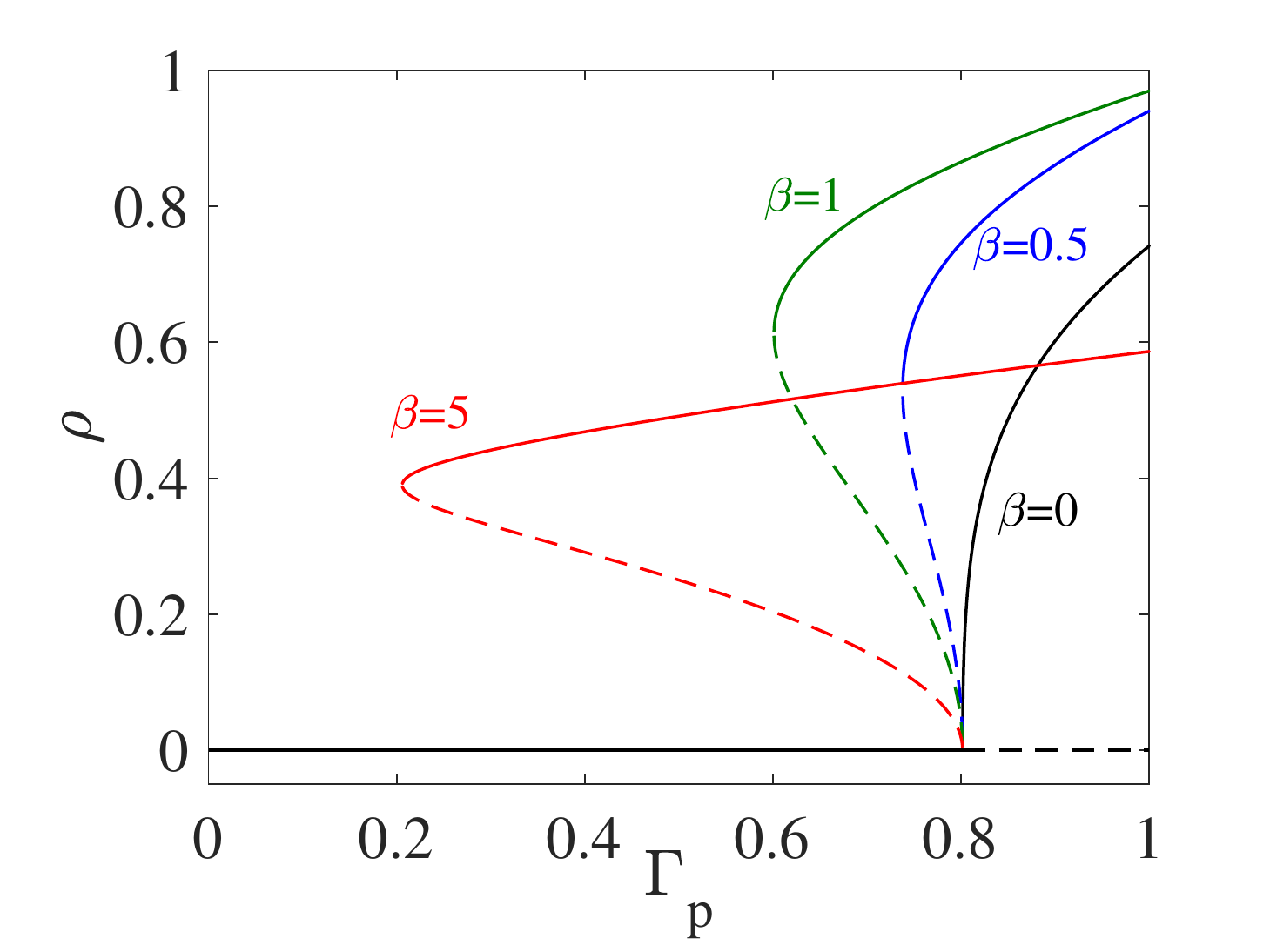}
		(c)\includegraphics[width=0.4\linewidth]{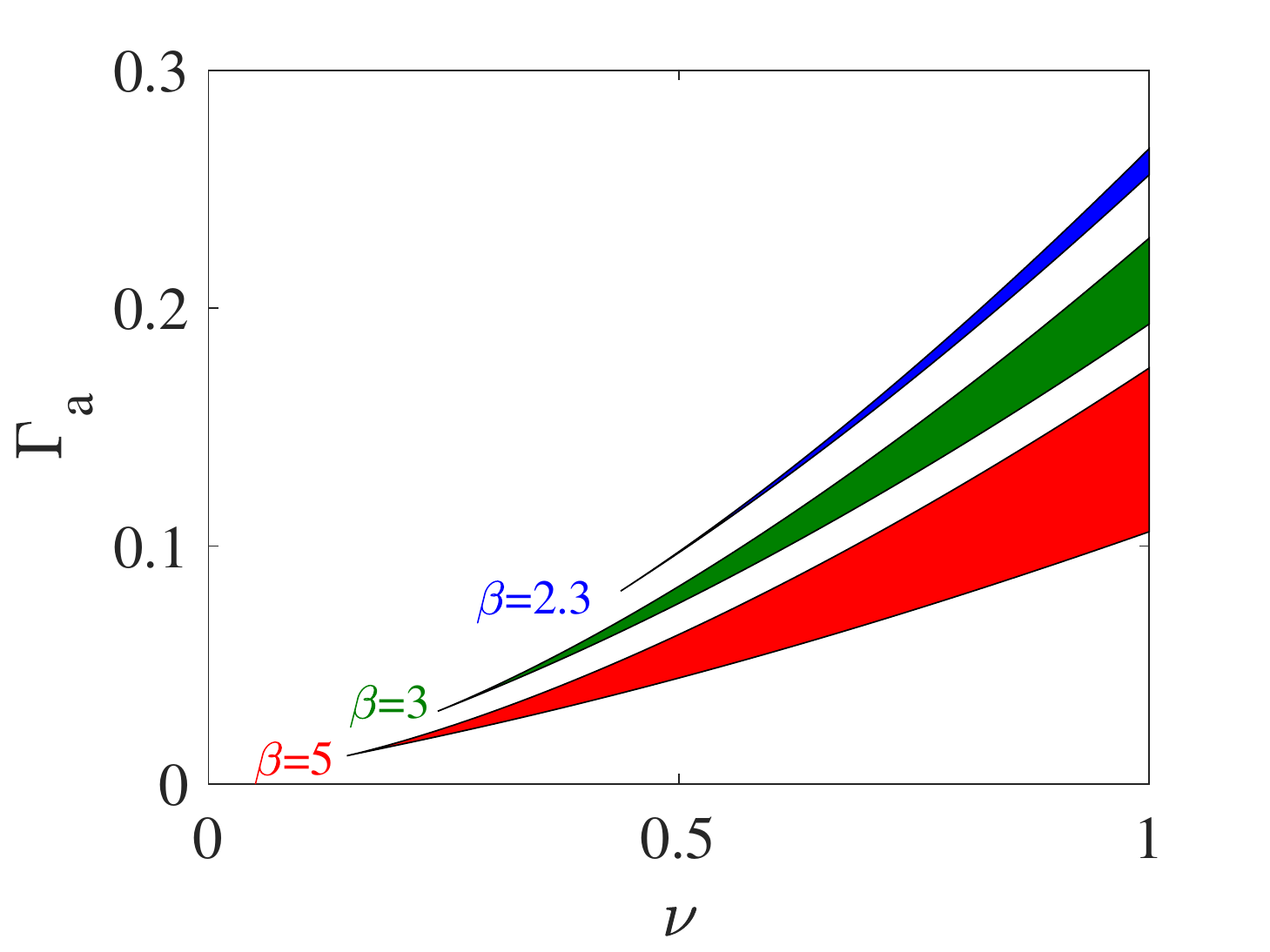}
		(d)\includegraphics[width=0.4\linewidth]{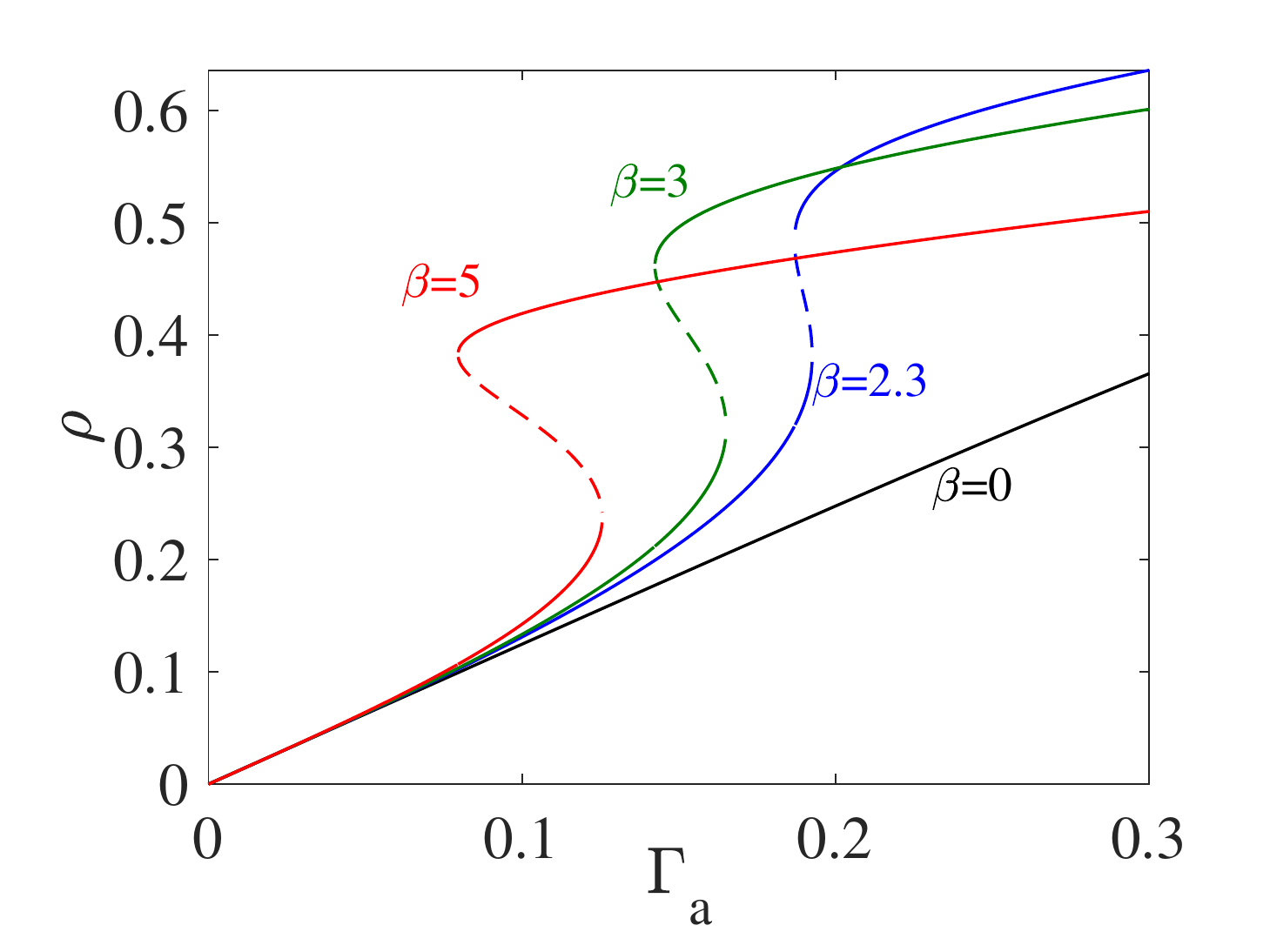}
		\caption{Bistability regions in the forcing parameter plane $(\nu,\Gamma)$ (left panels) and the corresponding bifurcation diagrams (right panels), for spatially uncoupled systems at different values of $\beta$. (a) Bistability regions of high-amplitude resonant oscillations and steady state (zero amplitude) in the case of pure parametric forcing according to~\eqref{pararegions}, and (b) the corresponding bifurcation diagrams according to~\eqref{eq:paraSOL} for $\nu=0.8$. (c) Bistability regions of high-amplitude and low-amplitude resonant oscillations in the case of pure additive forcing according to~\eqref{addbistregions}, and (d) the corresponding bifurcation diagrams according to~\eqref{eq:addSolutions} for $\nu=0.8$. Solid (dashed) lines in (b,d) indicate linearly stable (unstable) solutions (computed numerically).}
		\label{fig:bistabilityRegions}
	\end{figure}
	
	\subsubsection{Pure parametric forcing}
	
	Bistability is characterized by the coexistence of two linearly stable solutions of Eq.~\eqref{eq:FCGLmodel} in some parameter range. For pure parametric forcing, bistability is associated with a subcritical bifurcation of the zero state, $\rho=\rho_p^0$. The bistability range is given by $\Gamma_p^-<\Gamma_p<\Gamma_p^+$~\cite{BurkeYochelisKnobloch2008}, where
	\begin{equation}\label{pararegions}
	\Gamma_p^-=\dfrac{\nu-\beta\mu}{\sqrt{1+\beta^2}},\quad \quad
	\Gamma_p^+=\sqrt{\mu^2+\nu^2}.
	\end{equation}
	For damped oscillations $\mu<0$, the bistability range diminishes to zero at the codimension-2 point,
	\begin{equation}\label{gammaPCrit}
	\bra{\nu_c,\Gamma^c_p}=\bra{-\mu/\beta,-\mu\sqrt{1+1/\beta^2}},
	\end{equation}
	where the subcritical bifurcation of the zero state coincides with the saddle-node bifurcation where the two states $\rho_p^\pm$ merge and disappear. Note that
	$\Gamma^c_{p}\rightarrow |\mu|$ as $\beta\rightarrow \infty$ and $\nu\rightarrow 0$. Figure~\ref{fig:bistabilityRegions}(a) shows bistability regions for several values of $\beta$ in the parameter plane ($\nu,\Gamma_p$), where both the resonant 1:1 oscillations and the zero state are stable solutions, while Fig.~\ref{fig:bistabilityRegions}(b) shows the amplitude solutions at a fixed detuning value.
	
	\subsubsection{Pure additive forcing}
	
	In the case of a pure additive forcing, {a bistability range is found that involves low-amplitude and high-amplitude resonant 1:1 oscillations}, and is given by	$\Gamma^-_a<\Gamma_a<\Gamma^+_a$~\cite{yipingtheis}, where
	\begin{equation}\label{addbistregions}
	\Gamma_a^\pm=\sqrt{\frac{18(1+\beta^2 )(\mu+\beta\nu)(\mu^2+\nu^2 )-16(\mu+\beta\nu)^3\mp 2\left(4(\mu+\beta \nu)^2-3(1+\beta^2 )(\mu^2+\nu^2 )\right)^{3/2}}{27(1+\beta^2 )^2}},
	\end{equation}
	as Fig. \ref{fig:bistabilityRegions}2 shows for a few $\beta$ values. For $\beta>\beta_c=\sqrt{3}$, the range diminishes to zero in a cusp bifurcation at~\cite{yipingtheis}
	\begin{equation}\label{additiveCusp} \Gamma^{c}_a=-\frac{2\mu}{\sqrt{3}\left(\beta-\sqrt{3}\right)}\sqrt{\frac{-2\mu\bra{\beta^2+1}}{\sqrt{3}\left(\beta-\sqrt{3}\right)}}.
	\end{equation}
	Note that $\Gamma^{c}_a$ decreases as $\beta$ is increased (see Fig.~\ref{fig:bistabilityRegions}(b)), and that $\Gamma^{c}_a\rightarrow 0$ as $\beta\rightarrow \infty$. Figure~\ref{fig:bistabilityRegions}(c) shows bistability regions for several values of $\beta$ in the parameter plane ($\nu,\Gamma_a$), while Fig.~\ref{fig:bistabilityRegions}(d) shows the amplitude solutions at a fixed detuning value.
	
	\subsection{Combined parametric and additive forcing and tri-stability}
	
\begin{figure}[tp]
		(a)\includegraphics[width=0.31\linewidth]{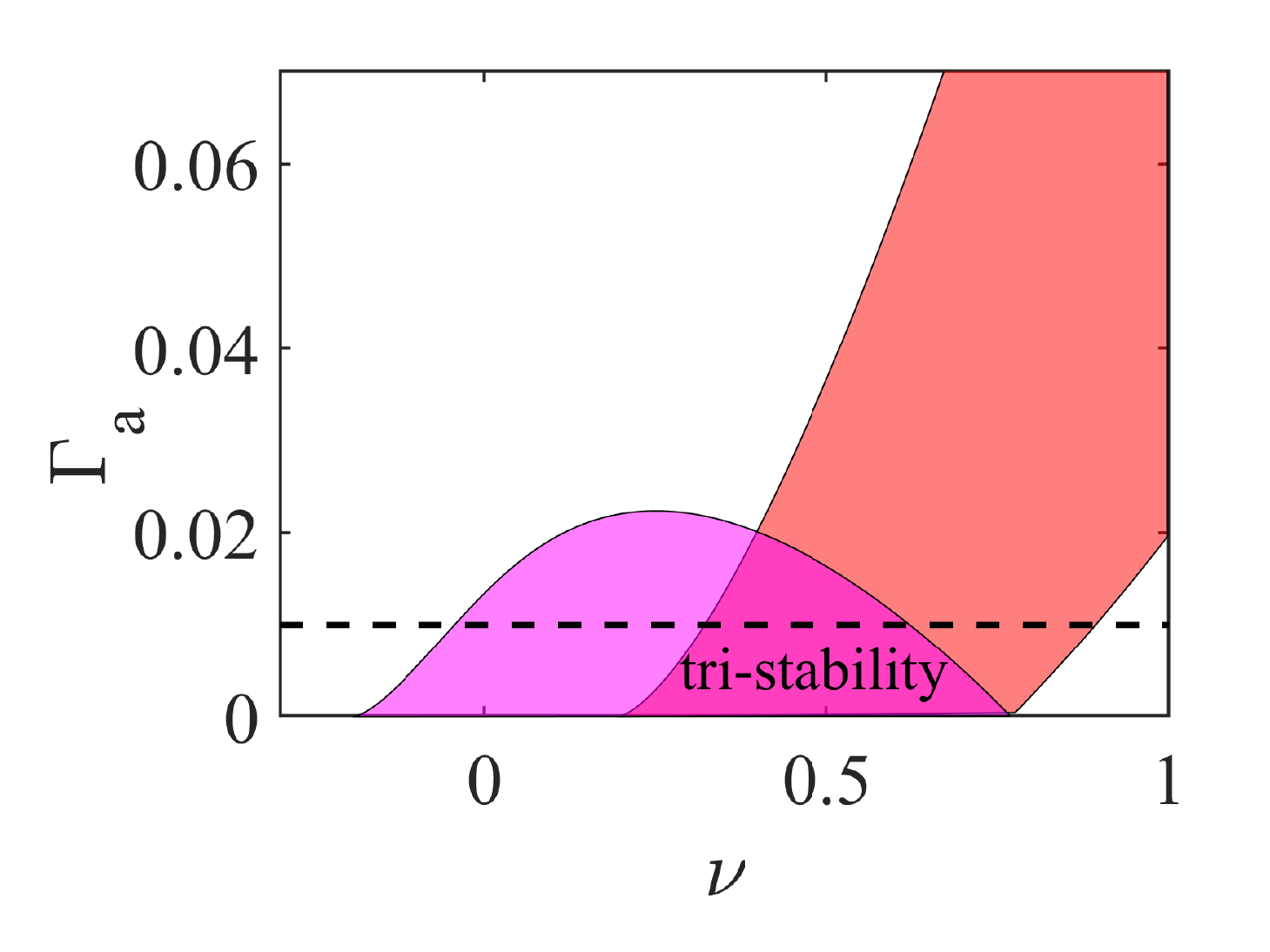}
		(c)\includegraphics[width=0.31\linewidth]{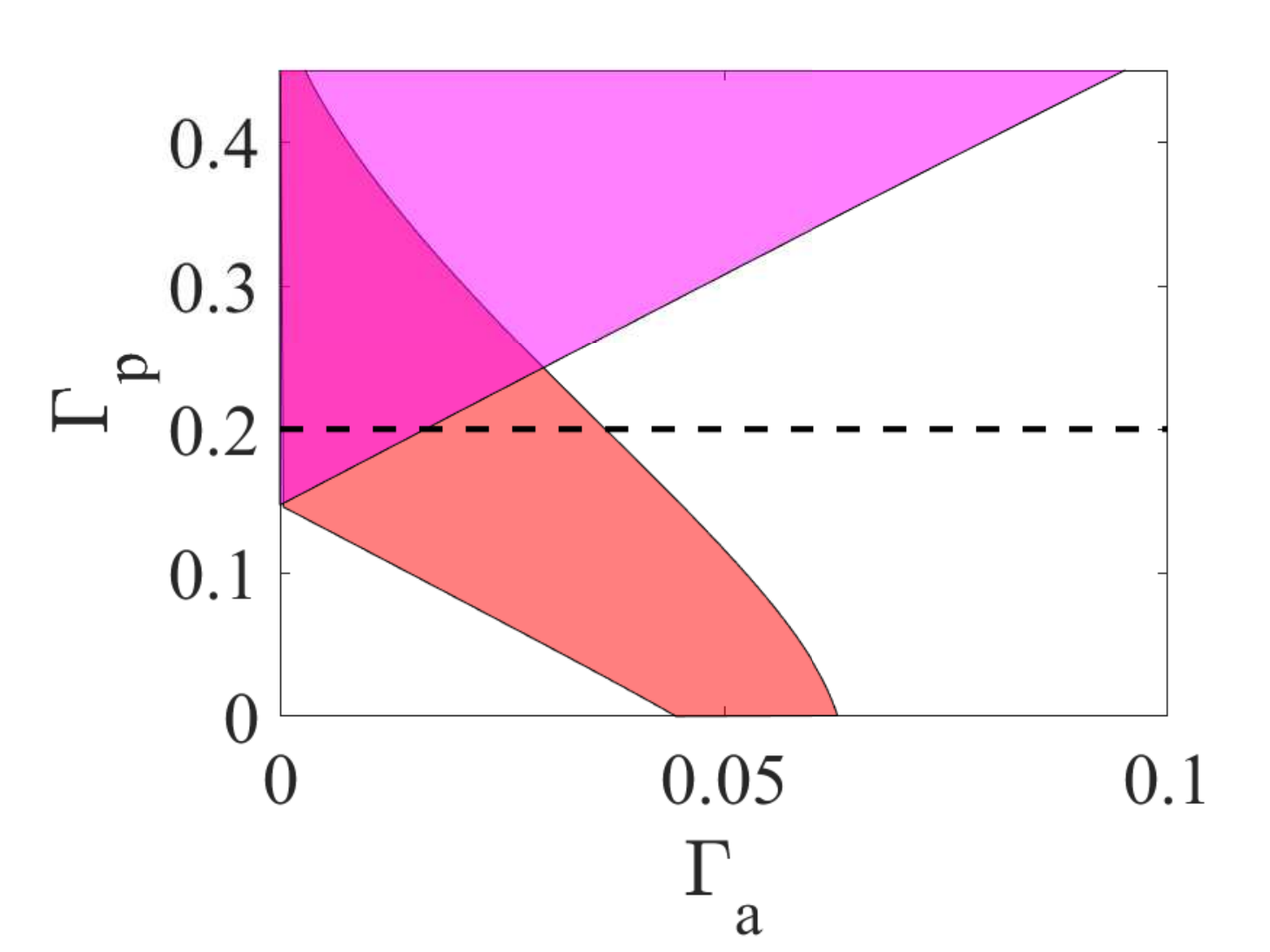}
		(e)\includegraphics[width=0.31\linewidth]{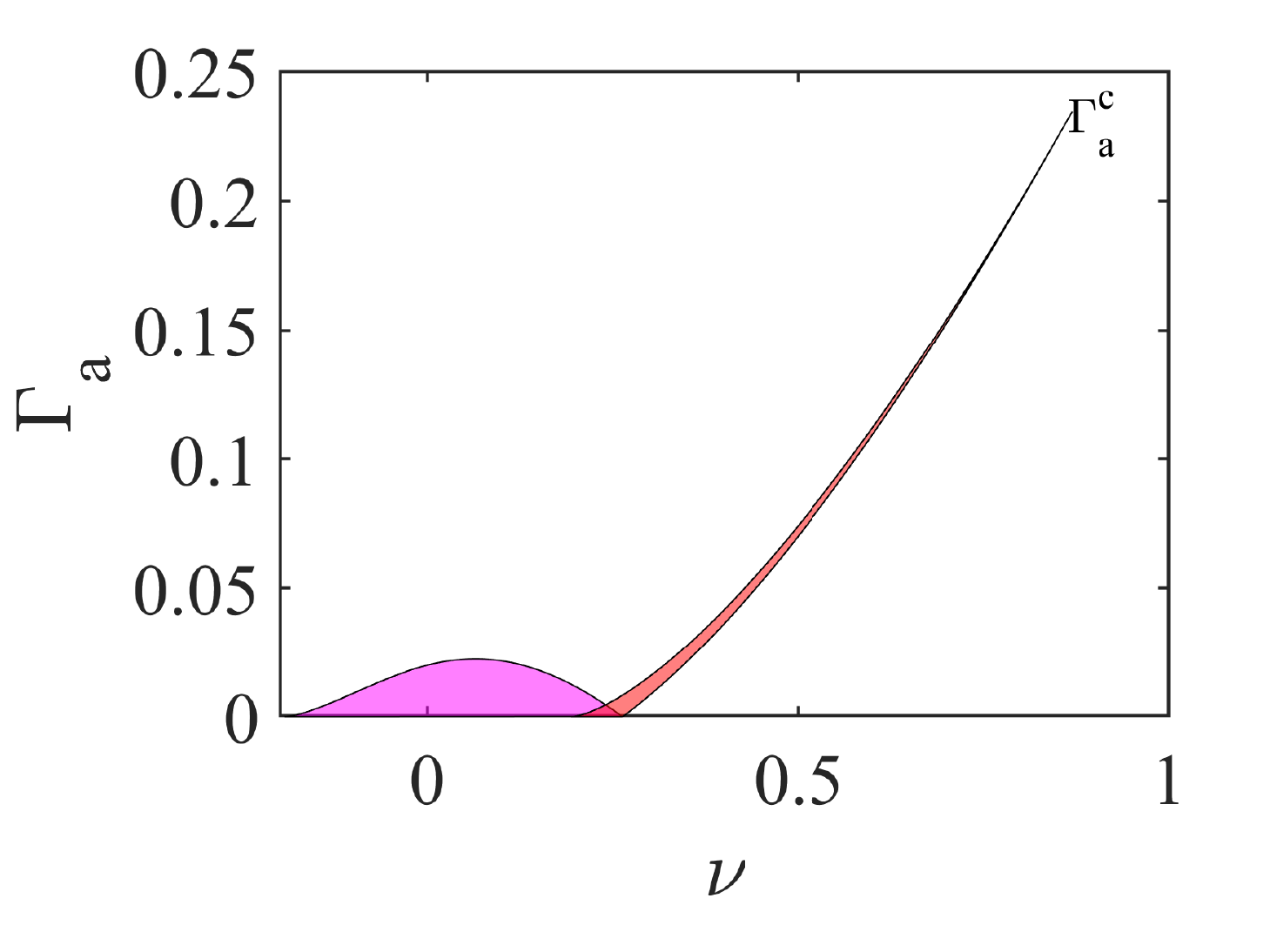}
		(b)\includegraphics[width=0.31\linewidth]{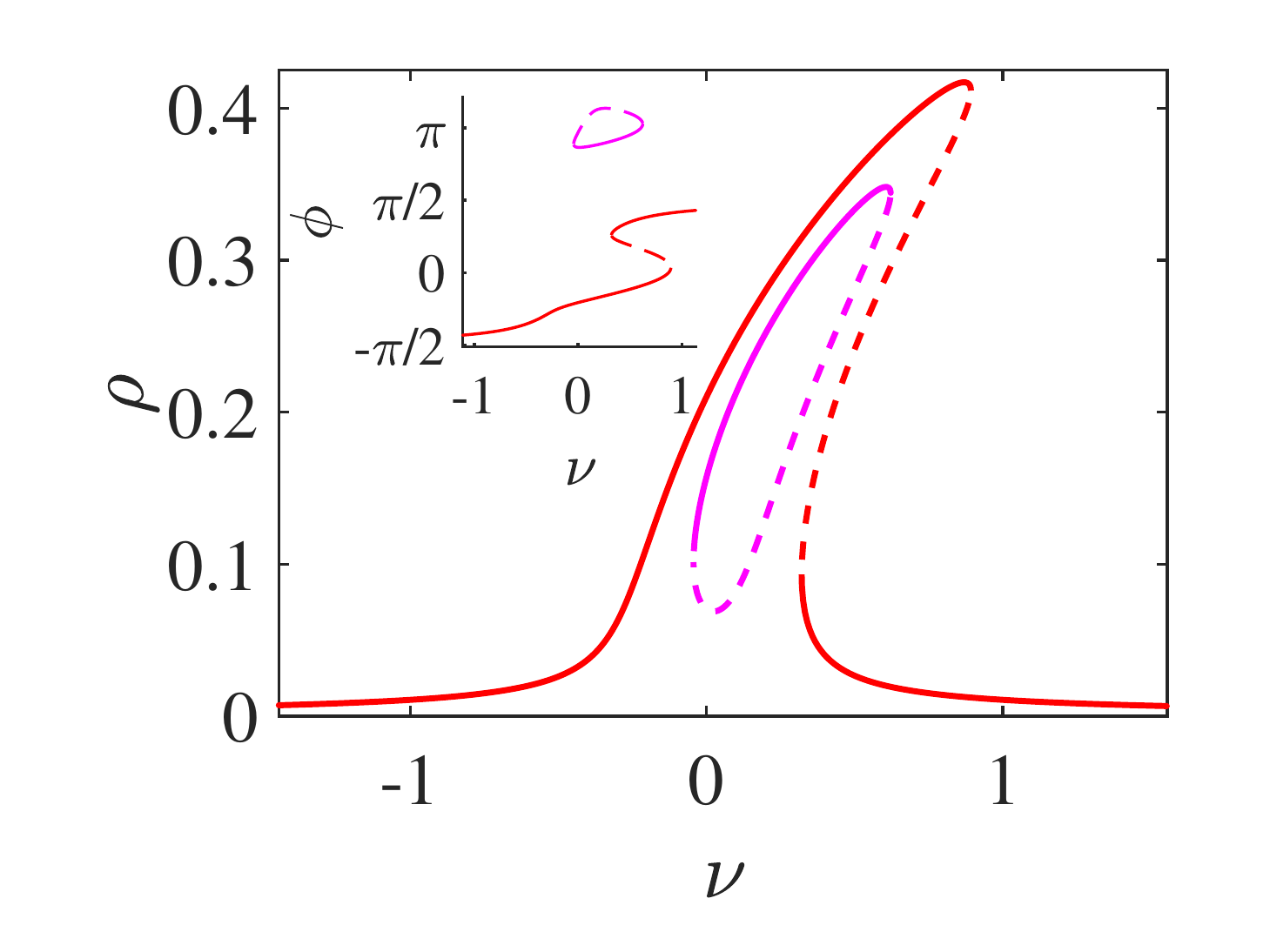}
		(d)\includegraphics[width=0.31\linewidth]{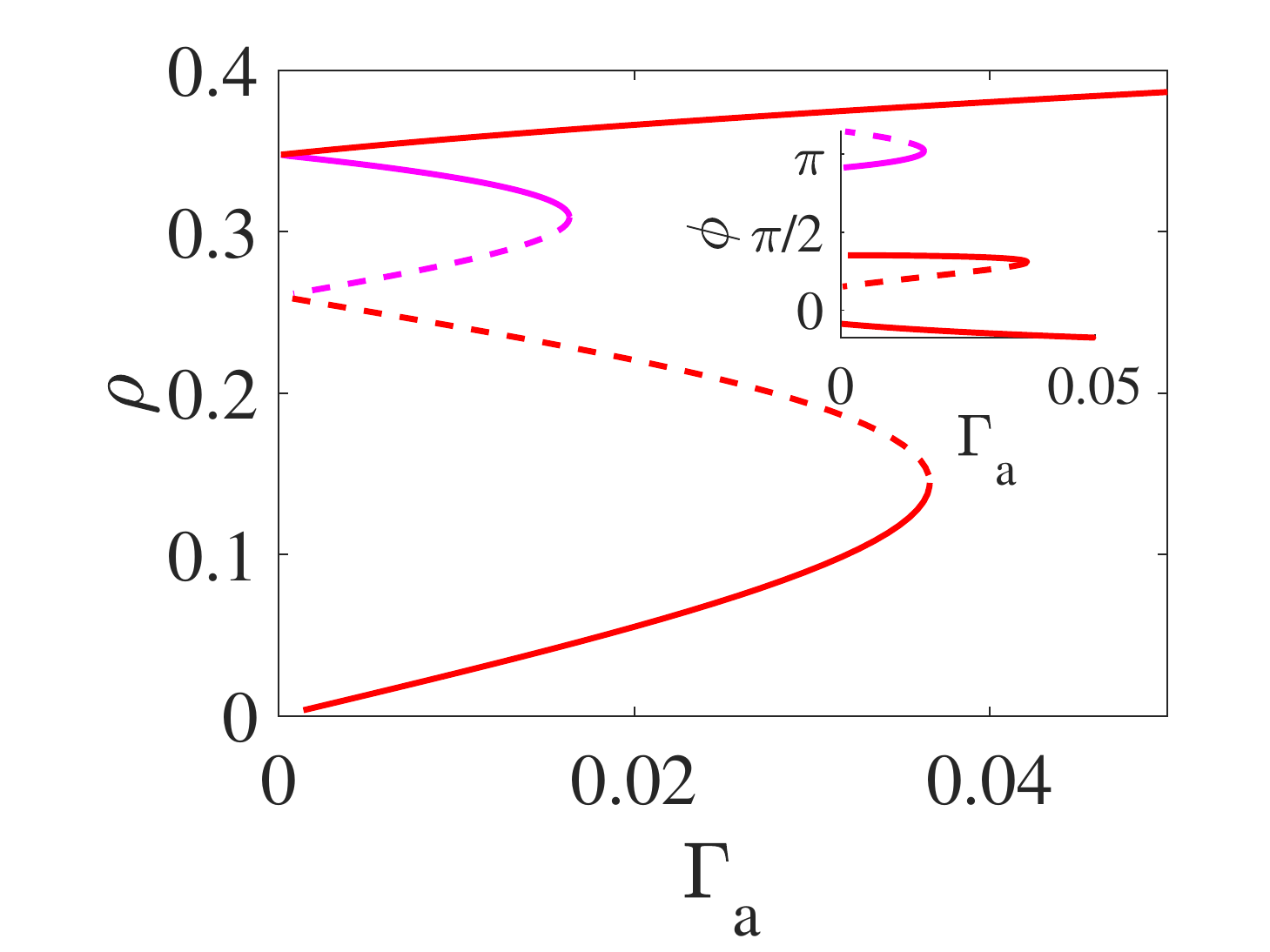}		
		(f)\includegraphics[width=0.31\linewidth]{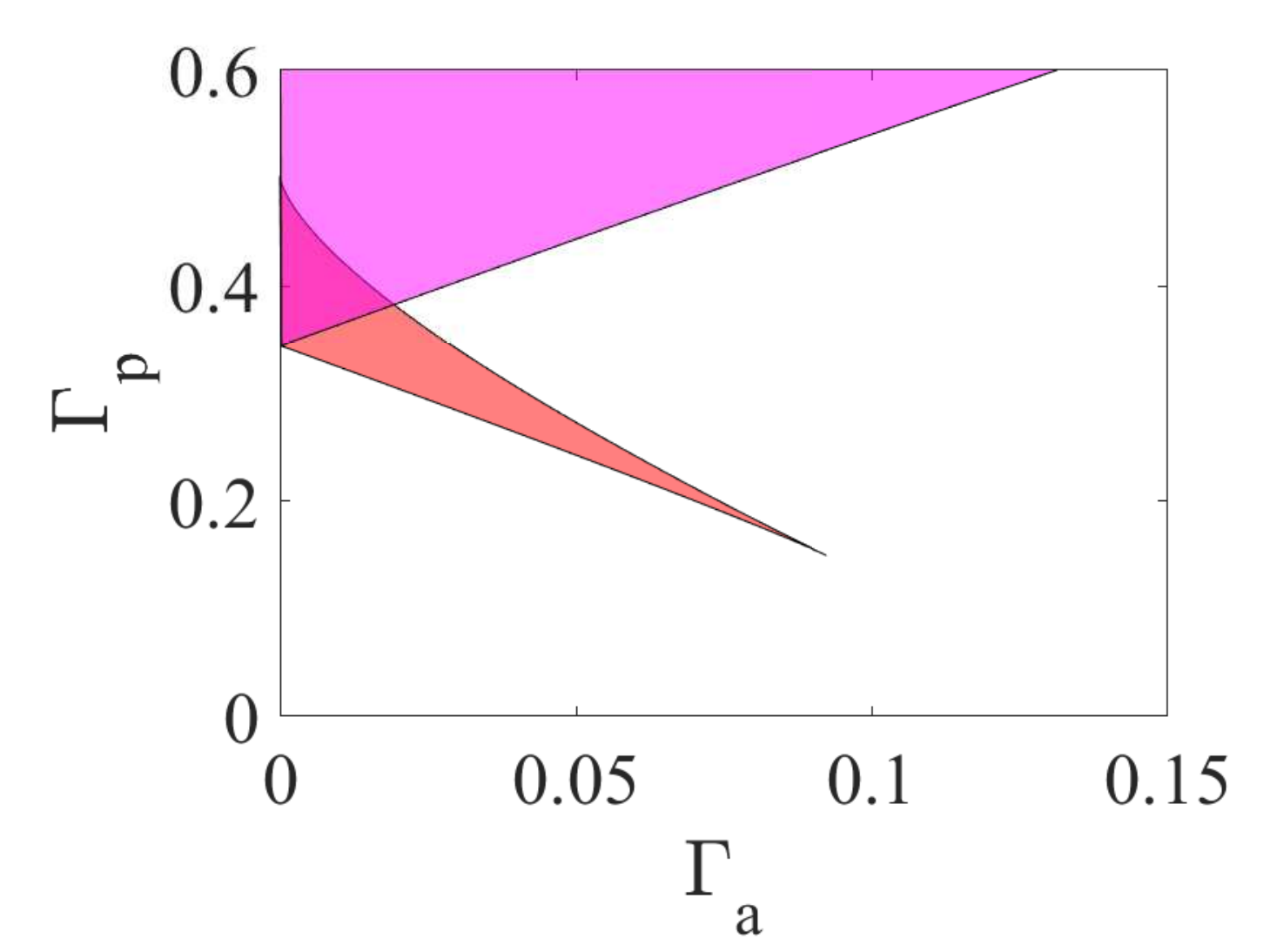}
		\caption{(a) Bistability and tristability regions in the ($\nu, \Gamma_a$) plane under combined forcing, computed numerically at $\beta=5$ and $\Gamma_p=0.2$. (b) Bifurcation diagrams in terms of $\rho(\nu)$ and $\phi(\nu)$ (see inset) at $\Gamma_a=0.01$ (dashed line in (a)). (c) Bistability and tristability regions in the ($\Gamma_a, \Gamma_p$) plane for $\nu=0.5$. (d) Bifurcation diagrams in terms of $\rho(\Gamma_a)$ and $\phi(\Gamma_a)$ (see inset)  at $\Gamma_p=0.2$. (e,f) Bistability and tristability regions for $\beta=1.3<\beta_c=\sqrt{3}$, showing inversion (in $\Gamma_a$) of the cusp bifurcation direction under combined forcing as compared to pure additive forcing. Solid (dashed) lines in (b,d) indicate stable (unstable) solutions.}
		\label{fig:combinedbistbeta5}
	\end{figure}

	Combined parametric and additive forcing is not amenable to analytic derivations, and thus, we used the numerical continuation package AUTO~\cite{doedel1998auto} to gain insights about the organization of coexisting solutions.
	The combination of the two forcing types can introduce an additional phase-shifted high-amplitude resonant solution, as Fig.~\ref{fig:combinedbistbeta5}(a,b) shows (purple shaded region in (a) and purple lines in (b)). These solutions appear for sufficiently high values of the parametric forcing, $\Gamma_p>\mu$ for which $\Gamma_a^c<0$, and give rise to a tristability parameter range in which all three resonant solution branches overlap: the low-amplitude and high-amplitude solutions that coexist as stable solutions already in the pure additive forcing case, and the additional phase-shifted high-amplitude solution (see overlap region in Fig.~\ref{fig:combinedbistbeta5}(a)). As the bifurcation diagrams in  Fig.~\ref{fig:combinedbistbeta5}(b) for the amplitude $\rho(\nu)$ and the phase $\phi(\nu)$ show, this additional solution branch organizes as an isola. A complementary representation of the three solutions and their existence and stability ranges is shown in Fig.~\ref{fig:combinedbistbeta5}(c,d). Surprisingly, a tristability range exists also for $\beta<\beta_c$, but the cusp bifurcation is in a reversed direction (Fig.~\ref{fig:combinedbistbeta5}(e)), although in the forcing parameter plane ($\Gamma_a,\Gamma_p$) the organization is qualitatively similar (Fig.~\ref{fig:combinedbistbeta5}(f)). Additionally, this cusp point vanishes, $\Gamma_a^c \to 0$, as $\beta \rightarrow \sqrt{(1-\Gamma_{p}^2\mu^2)^{-1}}$ (not shown here). 	
	
	\section{Mapping detuning to space: Asymmetry of spatially-decoupled {resonances} }\label{sec:profile_nu}
	
	The results obtained in the previous section for spatially uncoupled systems ($D=0$) can be used to obtain a zeroth-order approximation for the spatial behavior of weakly coupled systems ($D\ll 1$) by mapping them onto  the spatial axis through the relation $\nu(x)=2\eta x+\nu_0$. In order to derive the properties of localized profiles, such as those shown in Fig. \ref{fig:RhoVsD}, we will focus on the stable solution, $\rho(\nu)$, of \eqref{eq:stat_amplitude} of highest amplitude. The aim is to evaluate the asymmetry index of solutions of~\eqref{eq:FCGLmodel}, and the impact of coexisting solutions on that asymmetry for pure parametric and additive forcing.
	
	\subsection{Maximum amplitude value and location in space}
		We first calculate the maximum amplitude of the localized profile, $\rho_m$,  and the location of that peak, $x_m$, as illustrated in Fig.~\ref{fig:RhoVsD}. For $\beta=0$ and $\nu_0=0$ the peak is located at $x=0$ and the profile obeys the $x \to -x$ symmetry~\cite{ourChaos}. However, for $\beta\neq 0$, the location of the peak is shifted, and the peak becomes asymmetrical.
	
	The extremum of the amplitude, $\rho=\rho_m$ can be obtained by solving the equation:
	\begin{equation}\label{eq:lagrangemul}
	\frac{df}{d\nu}=\frac{\partial f}{\partial (\rho^2)}\frac{\partial (\rho^2)}{\partial \nu}+\frac{\partial f}{\partial \nu}=0,
	\end{equation}
	and since $\partial (\rho^2)/\partial \nu=0$ it follows that
	\begin{equation}\label{nuromax}
	\nu_{m}=\beta\rho_m^2.
	\end{equation}
	Evaluating the second derivative of \eqref{eq:lagrangemul} indicates that the extremum $\nu_m$ is indeed a maximum:
	\[
	\frac{\partial^2 (\rho^2)}{\partial \nu^2}{{\Big |}_{\rho=\rho_m}}=-\left(\frac{\partial^2 f}{\partial \nu^2}\right)/\left(\frac{\partial f}{\partial (\rho^2)}\right)=-\frac{2}{2(\rho^2-\mu)(\delta(1-\rho^2)+\rho^2)+(\delta-1)\Gamma_a^2/\rho^2}<0.
	\]
	Using~\eqref{nuromax} in~\eqref{eq:stat_amplitude}, we obtain the following amplitude maxima for pure additive or pure parametric forcing:
	\begin{equation}\label{eq:romax}
	\rho_m=\begin{cases}
	\sqrt{\Gamma_p+\mu}, & \quad \quad \text{for } \Gamma_p>0,\Gamma_a=0;\\\\
	\dfrac{1}{\sqrt{3}}\sqrt{2\mu+\mu^2\left(\dfrac{R_\gamma}{2}\right)^{-1/3}+\left(\dfrac{R_\gamma}{2}\right)^{1/3}}, & \quad \quad \text{for } \Gamma_p=0,\Gamma_a>0
	\end{cases}
	\end{equation}
	where $R_\gamma=27\Gamma_a^2-2\mu^3+3\sqrt{3}\Gamma_a\sqrt{27\Gamma_a^2-4\mu^3}$. This result generalizes a previous derivation for additive forcing by Egu\'{\i}luz \textit{et al.}~\cite{essential}, who assumed $\beta=0$, although it turns out that $\rho_m$ is independent of $\beta$. For additive forcing, the expression for the maximum amplitude $\rho_m$ shows a transition from linear to cubic root dependence on $\Gamma_a$, while for parametric forcing, it shows a square root dependence on $\Gamma_p$. These results extend earlier results obtained for $\beta=0$~\cite{ourPRE}.
	
	Finally, using the expression $\nu(x)=2\eta x+\nu_0$ for the spatial dependence of the detuning we obtain the peak location:
	\begin{equation}\label{eq:peakromax}
	x_m=\frac{\beta\rho_m^2}{2\eta}-\frac{\nu_0}{2\eta}\,.
	\end{equation}
	
	
	\subsection{Localization width and asymmetry}
	
	Another property of the localized profile that can be calculated, besides its maximum, $\rho_m$, and location, $x_m$, is its width, $W_x$ (see Fig.~\ref{fig:RhoVsD}). We first calculate the related width along the detuning axis, $W_\nu$, defined as $W_\nu=|\nu^{+}_{h}-\nu^{-}_{h}|$, where $\nu^{\pm}_{h}$ are the detuning values at which the amplitude $\rho$ drops down to half its maximum value, $\rho_m$, on the high and low detuning sides of the profile. Expressions for $\nu^{\pm}_{h}$ can readily be obtained using \eqref{eq:stat_amplitude} and \eqref{nuromax}:
	\begin{equation}\label{eq:nubi}	\nu^{\pm}_{h}=\frac{\nu_{m}}{4}\pm\sqrt{\bra{1-\delta}\left(\frac{\rho_m}{2}\right)^{-2}\Gamma_a^2+\delta\cdot \Gamma_p^2-\left(\left(\frac{\rho_m}{2}\right)^2-\mu\right)^2}.
	\end{equation}
	
\begin{figure}[tp]
		\centering
		(a)\includegraphics[width=0.4\linewidth]{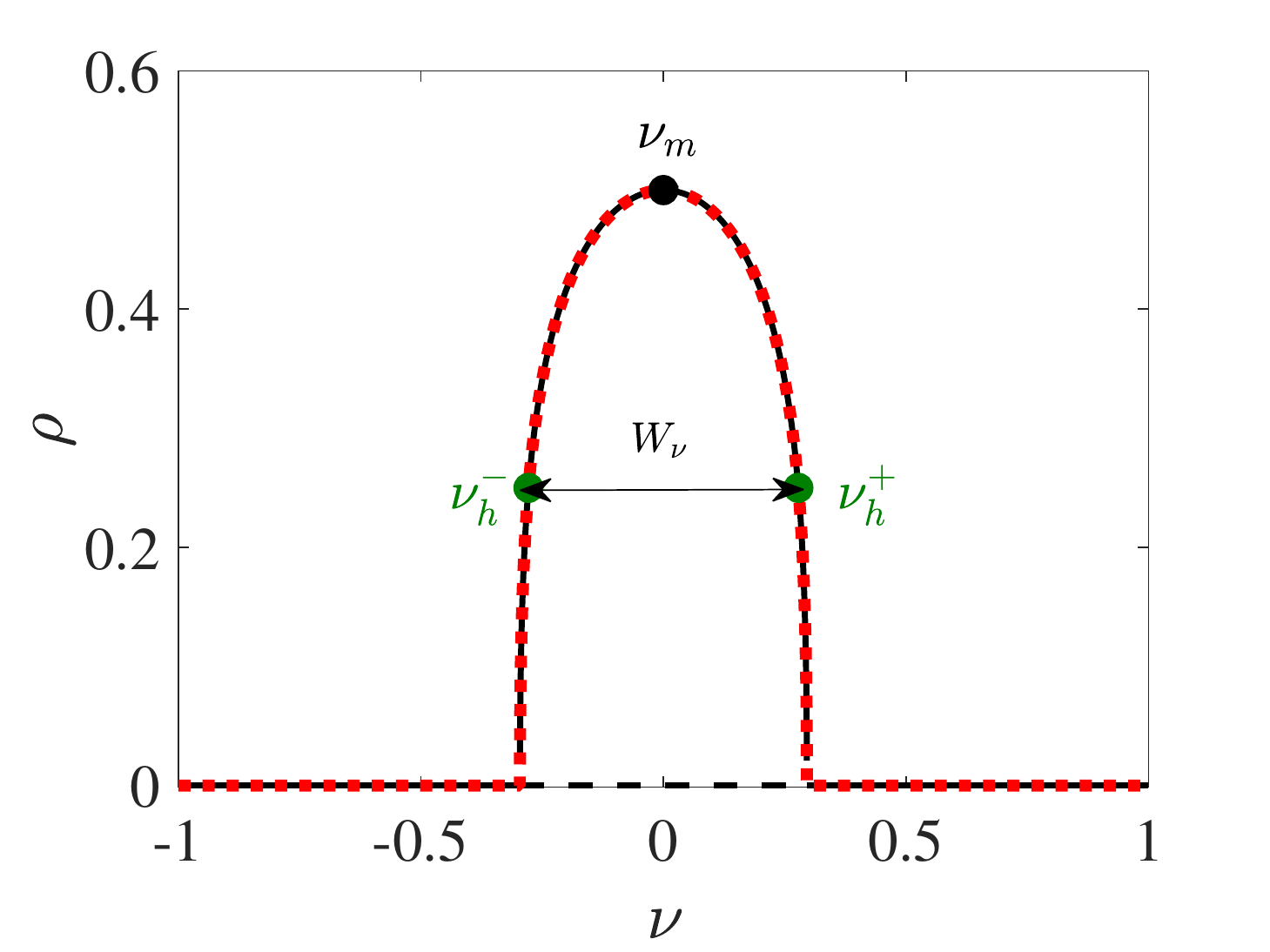}
		(b)\includegraphics[width=0.4\linewidth]{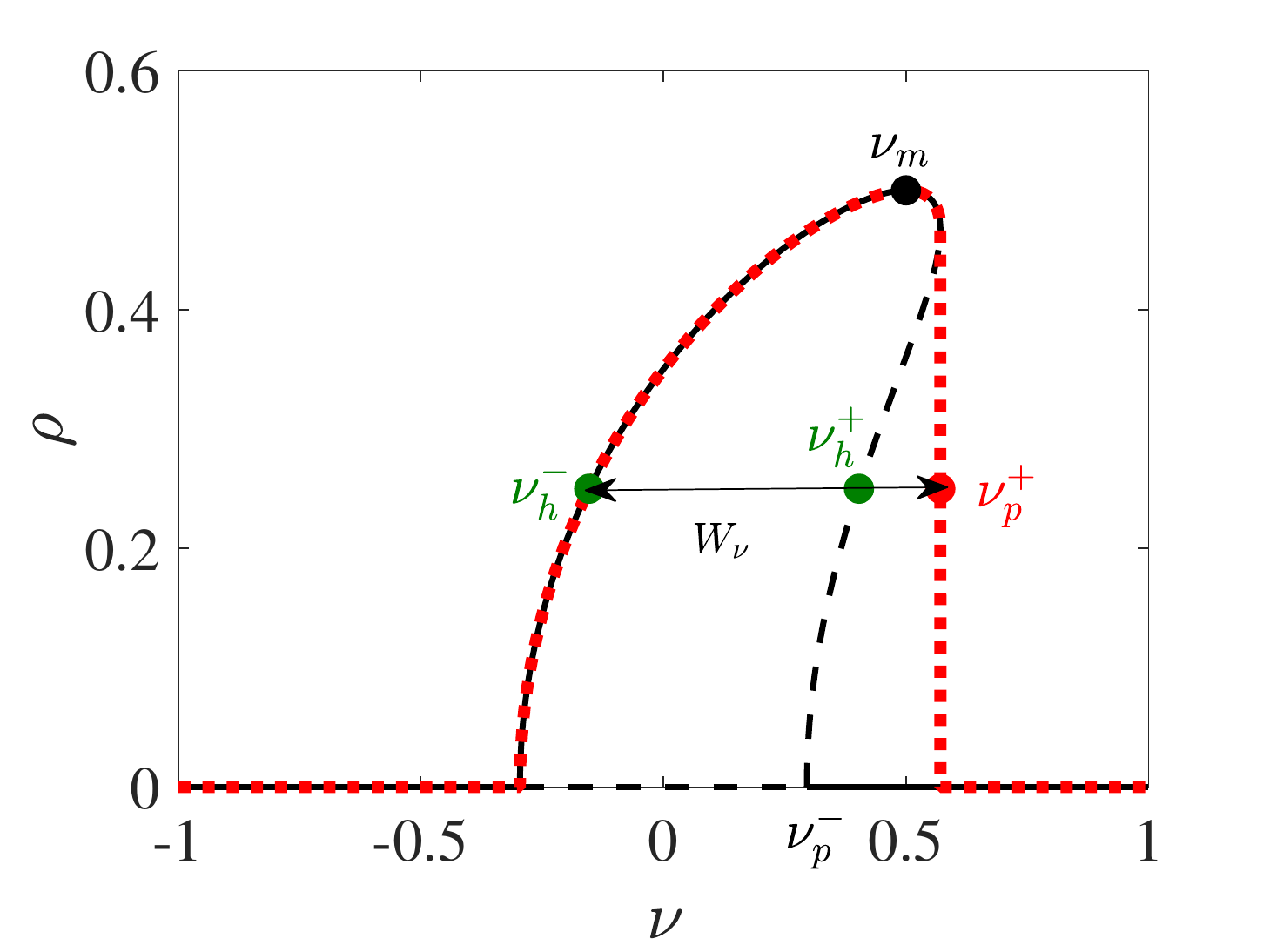}
		\caption{Spatial profiles of localized solutions illustrating the definition of the profile's width along the detuning axis, $W_\nu$, in the cases of (a) monostability ($\beta=0$) and (b) bistability ($\beta=2$); for both cases $D=0$. In case (a) the width is given by $|\nu^+_h - \nu^-_h|$ where $\nu^\pm_{h}$ stand for the detuning values at which the amplitude drops to one half of its the maximum value $\rho_m$ at $\nu=\nu_m$ ($\rho_m/2$). In case (b) the width is given by $|\nu^+_p - \nu^-_h|$, where $\nu^+_p$ is the detuning threshold of the saddle-node bifurcation that bounds the bistability range. The illustration is made here for the case of parametric forcing, but similar definitions hold for additive and combined forcing as well. Thin lines represent solutions to Eq.~\ref{eq:stat_amplitude} solid (dashed) lines designate stable (unstable) solutions, while dashed thick lines denote the profile of a localized resonant solution~\eqref{eq:FCGLmodel} (see text for details).}
		\label{fig:nuhalfexplain}
	\end{figure}

	In the absence of bistability (coexisting solutions), $W_\nu$ defines accurately the profile width, as shown in Fig.~\ref{fig:nuhalfexplain}(a). However, when one of $\nu^{\pm}_{h}$ falls inside the bistability region,
	that value should be replaced by the saddle-node value of $\nu$, e.g. $\nu^+_{p}$  as Fig.~\ref{fig:nuhalfexplain}(b) illustrates (recall that we are considering in this section spatially decoupled systems). Notably, although we conduct the analysis for the pure cases of parametric and additive forcing, the results also hold for combined forcing since the isola does not affect the width of the localized profile (see Fig.~\ref{fig:combinedbistbeta5}(b)).
	
	Consequently, for parametric forcing we obtain
	\begin{equation}\label{eq:widthpara}
	W_\nu=\begin{cases}
	\sqrt{\dfrac{15}{4}}\sqrt{(\Gamma_p+\mu)(\Gamma_p-3\mu/5)}, & \quad \begin{split}
	&\text{for $\beta=0$ and $\Gamma_p>-\mu$}\\
	&\text{or $\beta>0$ and $\nu^{+}_{h}<\nu^{-}_{p}$}
	\end{split};\\\\
	\Gamma_p\bra{\sqrt{1+\beta^2}-\dfrac{\beta}{4}}+\dfrac{3\beta\mu}{4}+\sqrt{\dfrac{15}{16}}\sqrt{(\Gamma_p+\mu)(\Gamma_p-3\mu/5)}, & \quad \text{for $\beta>0$ and $\nu^{-}_{p}<\nu^{+}_{h}<\nu^{+}_{p}$}
	\end{cases}
	\end{equation}
	where $\nu^{-}_{p}=\sqrt{\Gamma_{p}^2-\mu^2}$ and $\nu^{+}_{p}=\Gamma_p\sqrt{1+\beta^2}+\beta\mu$, while for additive forcing the width reads
	\begin{equation}\label{eq:widthadd}
	W_\nu=\begin{cases}
	\dfrac{4}{\rho_m}\sqrt{\Gamma_a^2-\left(\dfrac{\rho_m}{2}\right)^6+3\left(\dfrac{\rho_m}{2}\right)^4\mu-\left(\dfrac{\rho_m}{2}\right)^2\mu^2}, & \quad \begin{split}
	&\text{for $\beta=0$}\text{ or $\beta>0$ and $\nu^{+}_{h}<\nu^{-}_{a}$}
	\end{split}\\\\
	\nu^+_{a}-\dfrac{\beta \rho_m^2}{4}+\dfrac{2}{\rho_m}\sqrt{\Gamma_a^2-\left(\dfrac{\rho_m}{2}\right)^6+3\left(\dfrac{\rho_m}{2}\right)^4\mu-\left(\dfrac{\rho_m}{2}\right)^2\mu^2}, & \quad \text{for $\beta>0$ and $\nu^{-}_{a}<\nu^{+}_{h}<\nu^{+}_{a}$}
	\end{cases}
	\end{equation}
	where $\nu^{\pm}_{a}$ are solutions to~\eqref{addbistregions}. These results indicate that for $\nu^{+}_{h}<\nu^{-}_{a}$ there is no bistability and thus, the width is independent of $\beta$ (as also $\rho_m$).
	
	The profile width along the $x$ axis, $W_x$, is easily obtained from the results for $W_{\nu}$ through the relation
	\begin{equation}\label{eq:WxWnu}
	W_x=\frac{W_\nu}{2\eta}.
	\end{equation}
	
	Finally, we turn to the asymmetry measure $\Lambda$, given by~\eqref{eq:assymetryMeasure}, and calculate it using the equivalent form
	\begin{equation}\label{eq:Lambda_nu}
	\Lambda=\int_{\nu_m}^{\nu_{R}}\rho^2(\nu)d\nu\Big/\int_{\nu_{L}}^{\nu_{m}}\rho^2(\nu)d\nu.
	\end{equation}
	Notably, the choice of defining the width, $W_\nu$ at $\rho_m/2$ smears the difference between the parametric forcing and the additive forcing: since the analysis in both cases is performed in the proximity of saddle nodes the information of low amplitude stable solutions is not essential for the analysis. Hence we exploit, first, the analytic forms obtained for parametric forcing and then show numerically that a similar trend applies also for the additive forcing. Using~\eqref{nuromax} and~\eqref{eq:romax}, we obtain the integration limits in~\eqref{eq:Lambda_nu}:
	\begin{equation}\label{nu_max}
	\nu_{m}=\beta(\Gamma_p+\mu),
	\end{equation}
	\begin{equation}\label{assbound}
	\nu_L=\nu^{-}_{h}=\frac{\beta(\Gamma_p+\mu)-\sqrt{15}\sqrt{(\Gamma_p+\mu)(\Gamma_p-3\mu/5)}}{4},
	\end{equation}
	\begin{equation}
	\nu_R=\begin{cases}\nu^{+}_{h}=\dfrac{\beta(\Gamma_p+\mu)+\sqrt{15}\sqrt{(\Gamma_p+\mu)(\Gamma_p-3\mu/5)}}{4}, & \quad \begin{split}
	& \nu^{+}_{h}<\nu^{-}_{p}\\\\
	\end{split}\\\\
	\nu_{p}^+=\Gamma_p\sqrt{1+\beta^2}+\beta\mu, & \quad \nu^{-}_{p}<\nu^{+}_{h}<\nu^{+}_{p}
	\end{cases}.
	\end{equation}
	Calculating the integrals in~\eqref{eq:Lambda_nu}, we find
	\begin{equation}\label{eq:assymetry}
	\Lambda=\frac{2\mu\left(\nu_R-\nu_{m}\right)+\beta\left(\left(\nu_R\right)^2-\left(\nu_{m}\right)^2\right)-\Gamma_{p}^2(\beta+(1+\beta^2)\arctan\beta)
		+ g\left(\nu_R\right)  }{2\mu\left(\nu_{m}-\nu^-_{h}\right)+\beta\left(\left(\nu_{m}\right)^2-\left(\nu^-_{h}\right)^2\right)+\Gamma_{p}^2(\beta+(1+\beta^2)\arctan\beta)-g(\nu^-_{h})},
	\end{equation}
	where:
	\begin{equation*}
	g(z)=(z-\beta\mu)\cdot \sqrt{\Gamma_p^2(1+\beta^2)-(z-\beta\mu)^2}+(1+\beta^2)\Gamma_p^2\arctan\left(\frac{z-\beta\mu}{\sqrt{\Gamma_p^2(1+\beta^2)-(z-\beta\mu)^2}}\right).
	\end{equation*}
		
\begin{figure}[tp]
		\centering
		(a)\includegraphics[width=0.4\linewidth]{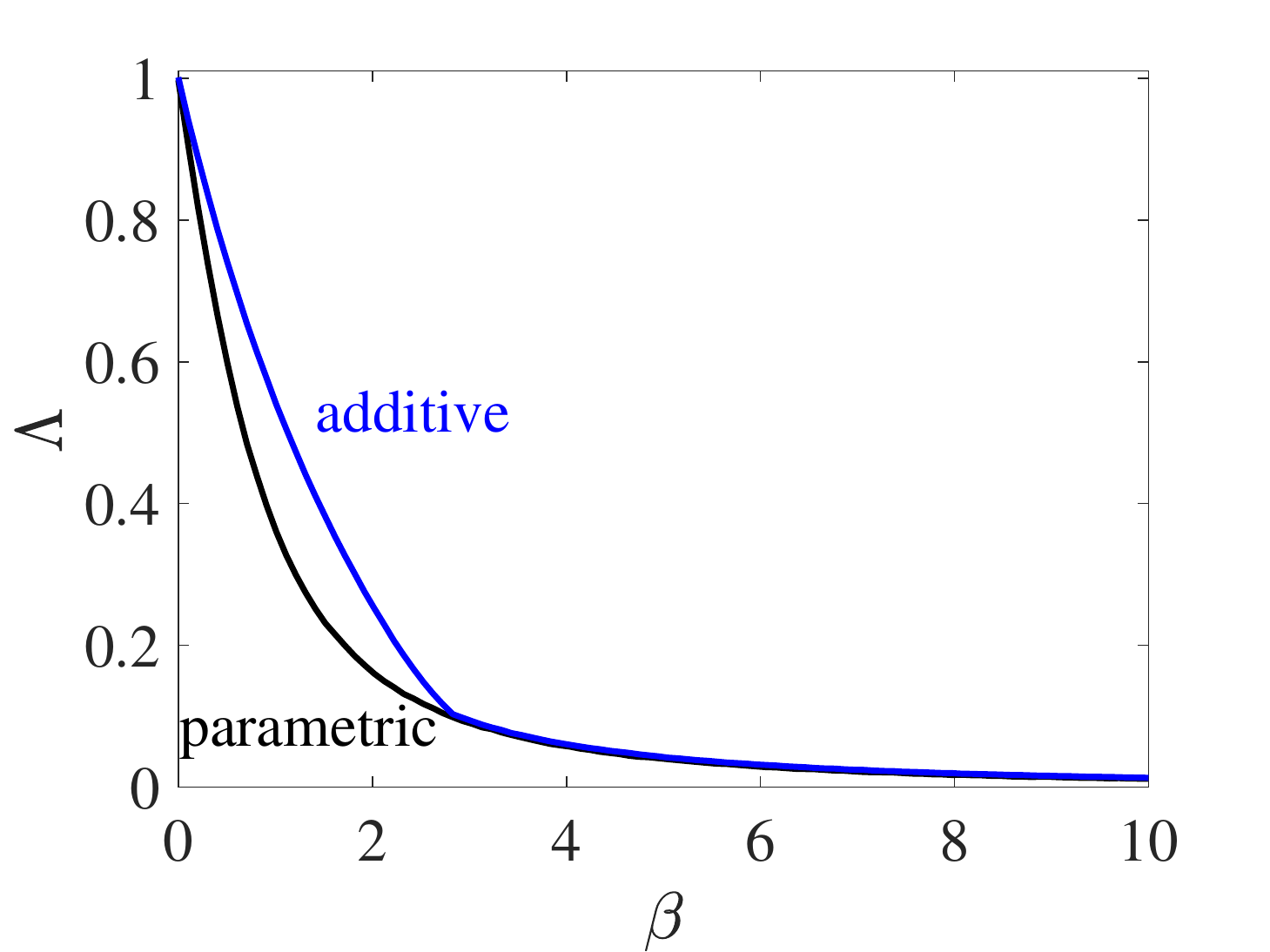}
		(b)\includegraphics[width=0.4\linewidth]{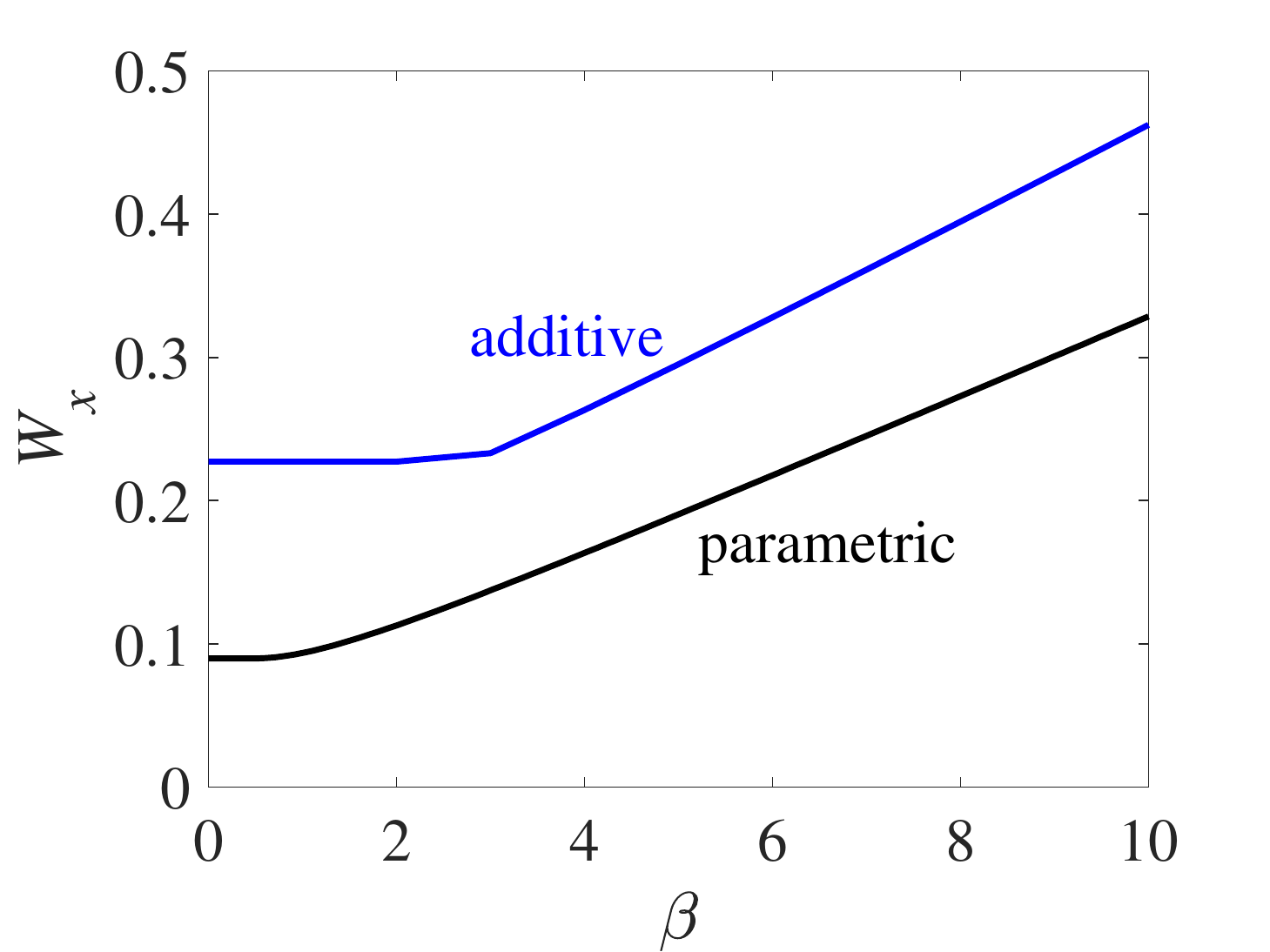}
		\caption{(a) Asymmetry index $\Lambda$ and (b) width of the spatial profile $W_x$ as a function of $\beta$ for the uncoupled case ($D=0$). The asymmetry index for the parametric forcing is obtained through~\eqref{eq:assymetry} while the width through~\eqref{eq:widthpara} and \eqref{eq:WxWnu}. For the additive case, the asymmetry is computed numerically from the profile $\rho(x)=\rho(\nu(x))$, where $\rho(\nu)$ is taken to be the largest solution of \eqref{eq:addSolutions}. The width in the additive case is obtained via~\eqref{eq:widthadd}. For the parametric case $\Gamma_p=0.2$, $\Gamma_a=0$, and for the additive case $\Gamma_p=0$, $\Gamma_a=0.1$.}
		\label{fig:profilesvsbetaPara}
	\end{figure}

	The dependencies of $W_x$ and of $\Lambda$ on $\beta$ are shown in Fig. ~\ref{fig:profilesvsbetaPara}. Notably, $\Lambda=1$ for $\beta=0$, as expected for a symmetric profile. As $\beta$ increases, asymmetry develops and $\Lambda$ decreases. The width of the spatial profile remains fairly constant at sufficiently small $\beta$ values, but beyond a threshold $\beta$ value at which subcriticality develops (see Fig.~\ref{fig:nuhalfexplain}), the width increases with $\beta$.
		
	\section{Resonant localization in spatially coupled systems}\label{sec:spatial}
	
\begin{figure}[tp]
		\centering
		(a)\includegraphics[width=0.3\linewidth]{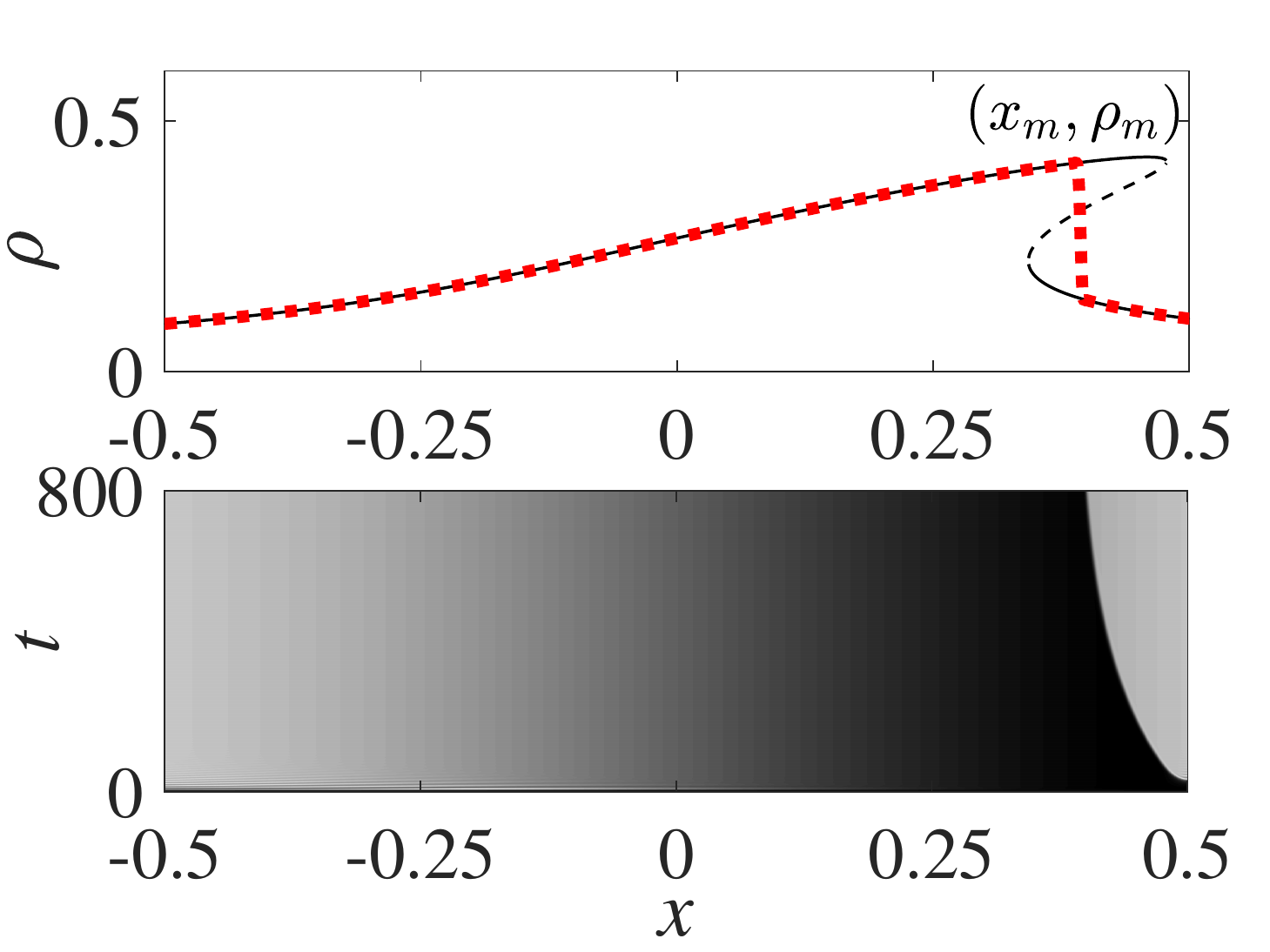}
		(b)\includegraphics[width=0.3\linewidth]{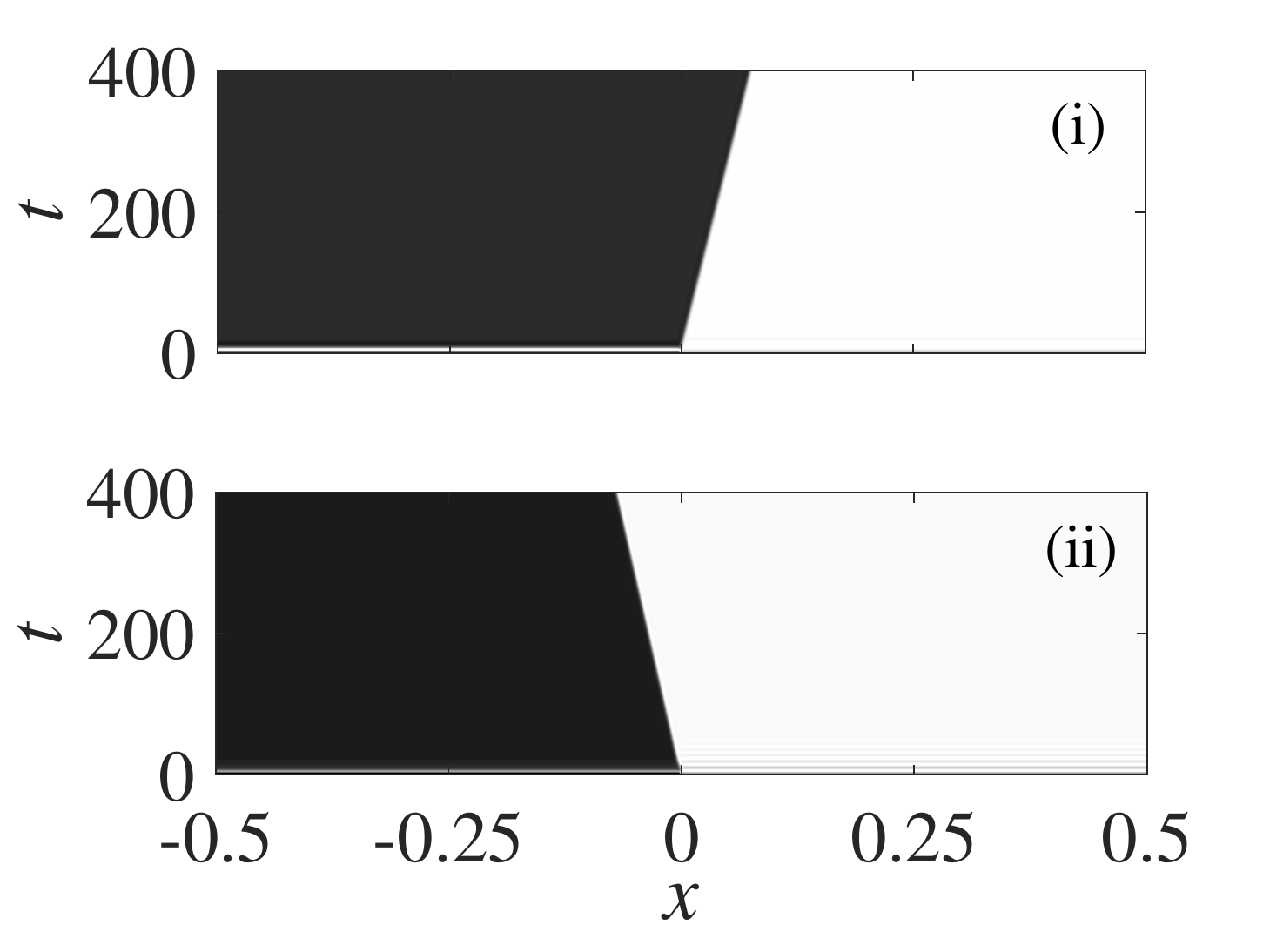}
		(c)\includegraphics[width=0.3\linewidth]{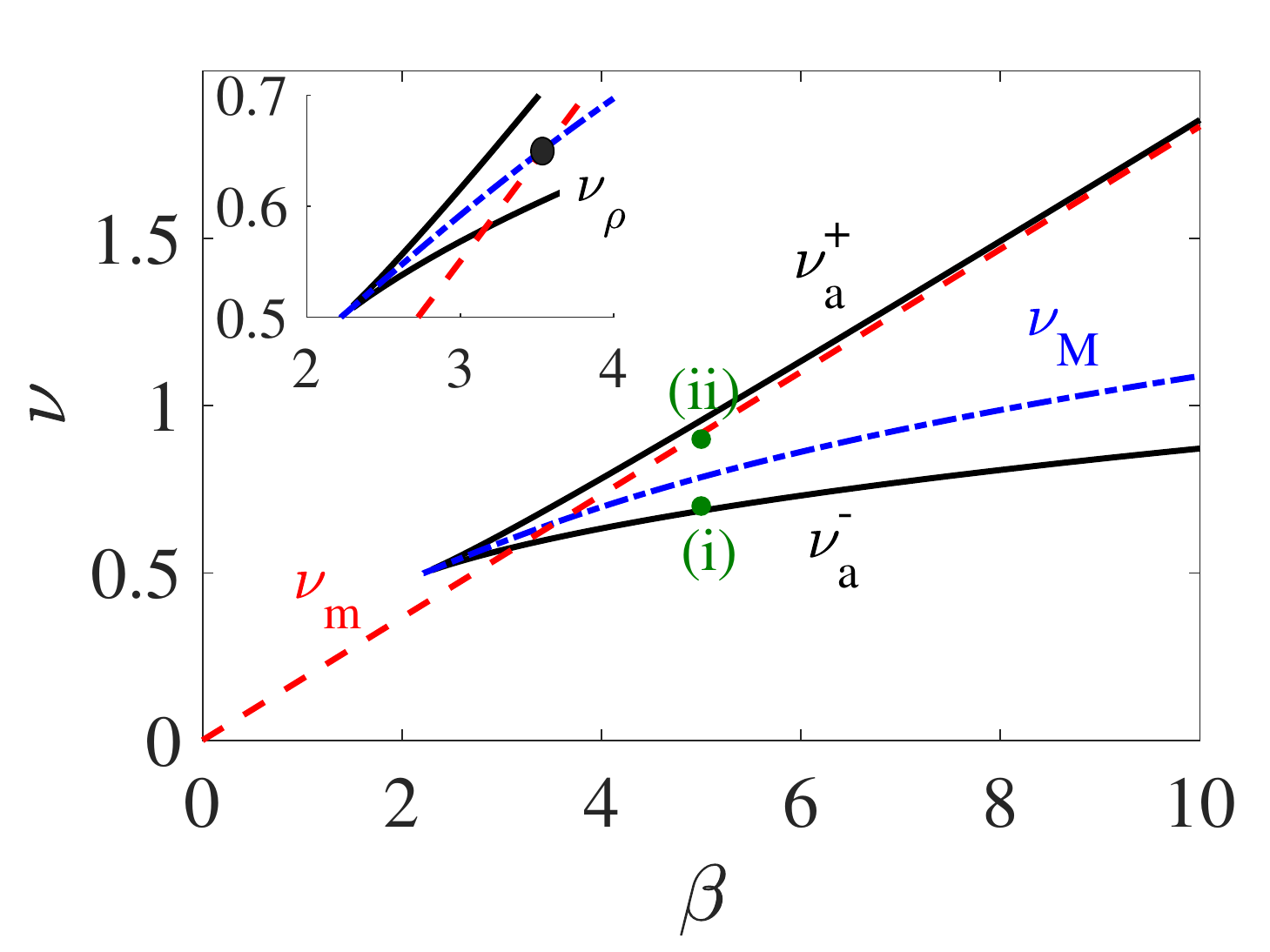}
		\caption{Asymmetric localized solutions and their Maxwell-point approximations. (a) A space-time plot (bottom panel) showing the development of a stationary localized solution of~\eqref{eq:FCGLmodel} with a sharp front (darker shades indicate larger amplitude values), and the spatial profile of that solution (dotted line in top panel). The thin solid and dashed lines represent solution $\rho(\nu)$ of the uncoupled system.
			(b) Space-time plots for a uniform system ($\eta=0$), showing fronts dynamics (i) below the Maxwell point $\nu_M$ ($\nu_0=0.7$) and (ii) above that point ($\nu_0=0.9$), as indicated in (c).	
			(c) Parameter plane ($\beta, \nu$) indicating the bistability range for additive forcing ($\nu^\pm_a$, solid lines), the location of the maximum amplitude ($\nu_m$, dashed line), and the location of the Maxwell point ($\nu_M$, dashed-dotted line) at which fronts in the homogeneous system are stationary. The inset shows a zoom in near the cusp bifurcation and indicates the $\nu_\rho$ point at which $\nu_m$ and $\nu_M$ intersect. Other parameters: $\Gamma_a=0.1$, $\Gamma_p=0$, $\eta=1$, $D=10^{-6}$, $\beta=5$.}
		\label{fig:weakcoupling}
	\end{figure}

	The spatial profile of localized oscillations in the bistability range, implied by Fig. \ref{fig:nuhalfexplain}(b), changes significantly when weak spatial coupling, $D\ll 1$, is added. The sharp front does not occur at the saddle-node point ($\nu_p^+$ in Fig.  \ref{fig:nuhalfexplain}(b)) but rather earlier. Direct numerical integration of \eqref{eq:FCGLmodel} indicates that the profile follows the upper branch, but then drops to the lower branch before reaching the right-most saddle-node, as Fig.~\ref{fig:weakcoupling}(a) shows. Since asymptotically the front resides at a specific location, its position may be approximated by considering a stationary front solution of the homogeneous system for which $\nu$ is constant. {In general}, front solutions of the bistable homogeneous system propagate, as Fig.~\ref{fig:weakcoupling}(b) shows, but there exists a particular $\nu$ value at which the front is stationary, the so-called Maxwell point, denoted here as $\nu_M$~\cite{YochelisBurkeKnobloch2006}. The front location is then given by $x_M=(\nu_M-\nu_0)/2\eta$.
	Since~\eqref{eq:FCGL} is not gradient~\cite{Lega1990}, we compute $\nu_M$, and its dependence on the parameter $\beta$ that controls the profile's asymmetry, by numerical continuation~\cite{doedel1998auto}. Figure~\ref{fig:weakcoupling}(c) shows a graph of $\nu_M$ as a function of $\beta$ (dashed-dotted line), which lies within the bistability range bounded by the solid lines $\nu=\nu_a^\pm(\beta)$. Also shown is the dependence of  $\nu_m$ on $\beta$ (dashed line), where $\nu_m$ is the $\nu$ value of the largest amplitude for the uncoupled case ($D=0$).
	
\begin{figure}[tp]
		\centering
		\includegraphics[width=0.8\linewidth]{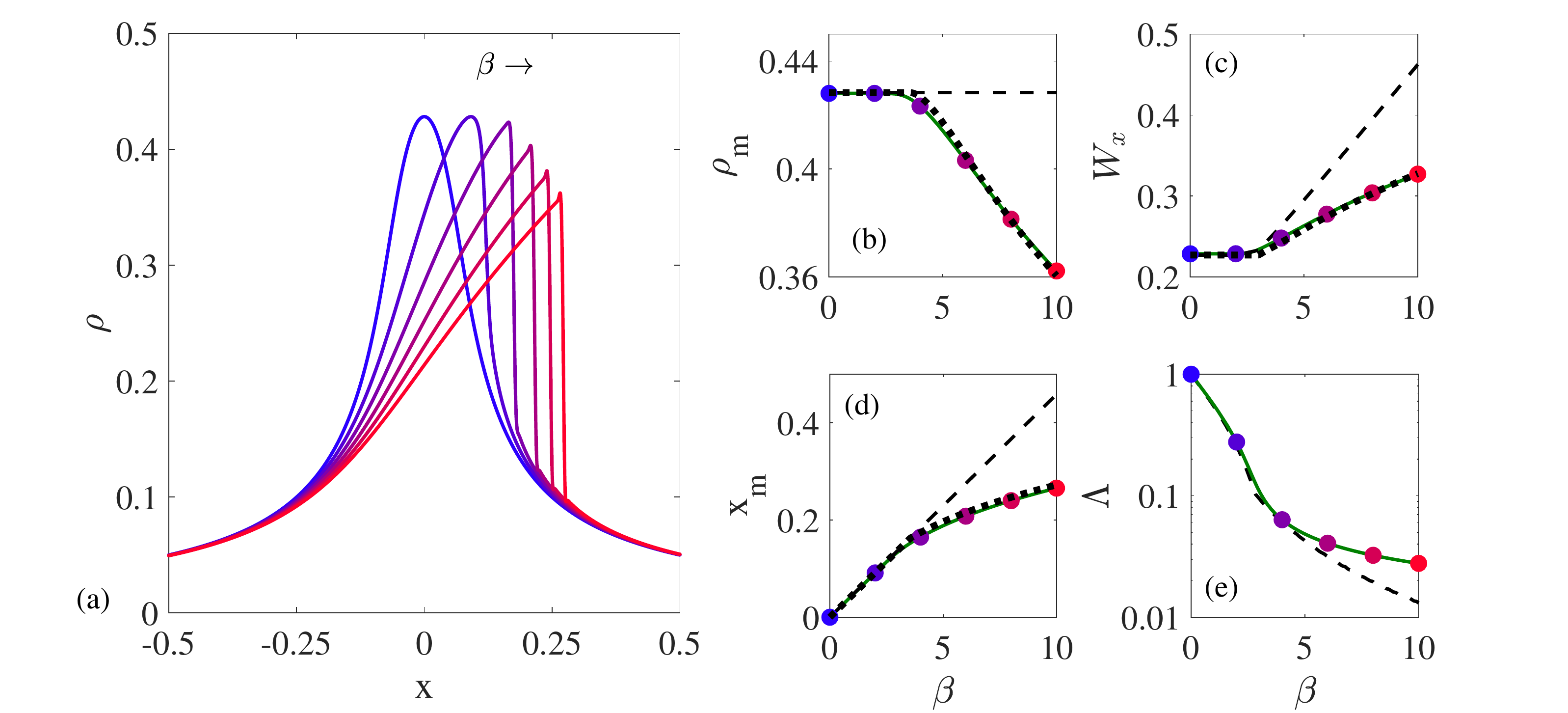}
		\caption{{Profile properties for additive forcing ($\Gamma_a=0.1,\Gamma_p=0$). (a) Amplitudes ($\rho$) computed by direct numerical integration of~\eqref{eq:FCGLmodel} for $\beta=0, 2, 4, 6, 8, 10$; the arrow indicates the increasing direction of $\beta$ values. (b-e) Maximum amplitude value ($\rho_m$), the amplitude peak location in space ($x_m$), profile width ($W_x$), and asymmetry index ($\Lambda$), respectively. Solid lines indicate results obtained by numerical integration of~\eqref{eq:FCGLmodel} for a weakly coupled medium ($D=10^{-6}$), while dotted lines represent results obtained by Maxwell-point calculations via numerical continuation of stationary fronts and accounting for the relation between $\rho(x)$ and $\rho(\nu)$, such that for $\nu<\nu_{M}$ the amplitude is $\rho=\rho_a^+$ (see~\eqref{eq:addSolutions_b}) and and for $\nu>\nu_{M}$ the amplitude $\rho=\rho_a^0$ (see~\eqref{eq:addSolutions_a}), and dashed lines indicate results obtained for an uncoupled medium ($D=0$), where the bistability region ends at $\nu=\nu_{p}^+$ (see Fig.~\ref{fig:nuhalfexplain}). The circles in (b-e), correspond to profiles in (a) at the respective values of $\beta$.}}
		\label{fig:weakCouplingProfilesVsBeta}
	\end{figure}

	The intersection point of the curves $\nu=\nu_M(\beta)$ and $\nu=\nu_m(\beta)$ (see Fig.~\ref{fig:weakcoupling}(c)) denotes the $\beta$ value beyond which the effect of spatial coupling becomes significant, as shown in 		Figure~\ref{fig:weakCouplingProfilesVsBeta} for $\rho_m$ (maximum amplitude), $x_m$ (location of maximum amplitude), $W_x$ (profile's width), and $\Lambda$ (asymmetry). Note the excellent agreement between results obtained by direct numerical integration of~\eqref{eq:FCGLmodel} (solid lines in panels (b-d) and the Maxwell-point approximations (dotted lines) obtained via continuation of stationary front solutions. Numerical computations demonstrate that the asymmetry index $\Lambda$ increases with $\beta$ in comparison with the decoupled case $D=0$; namely, the profile becomes more symmetric, see Fig.~\ref{fig:weakCouplingProfilesVsBeta}(e). We stress that for larger values of $D$ the symmetry is fully restored due to synchronization between strongly coupled oscillations~\cite{ourChaos}, see also Fig.~\ref{fig:RhoVsD}.
	
\begin{figure}[tp]
		\centering
		(a)\includegraphics[width=0.4\linewidth]{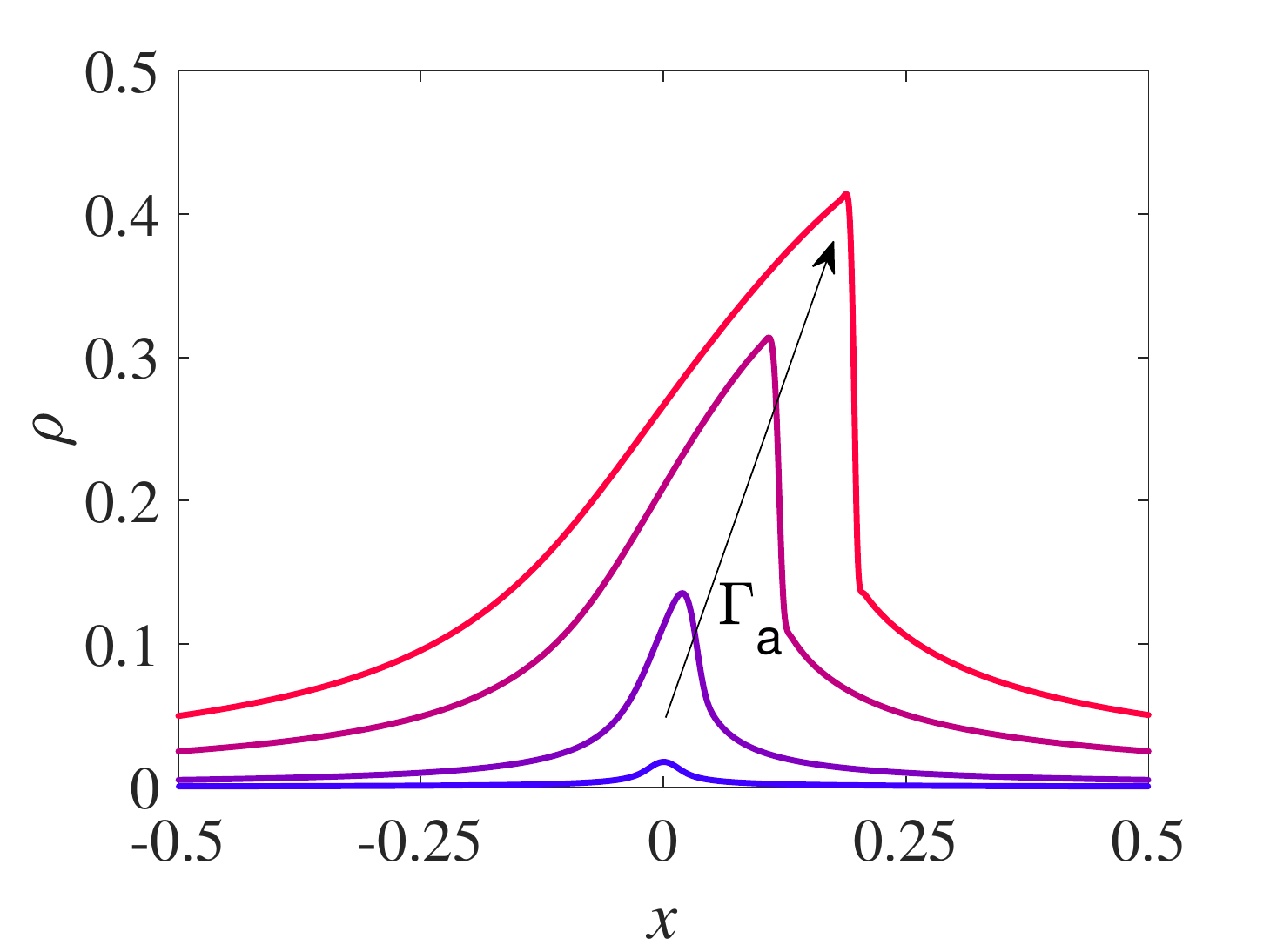}
		(b)\includegraphics[width=0.4\linewidth]{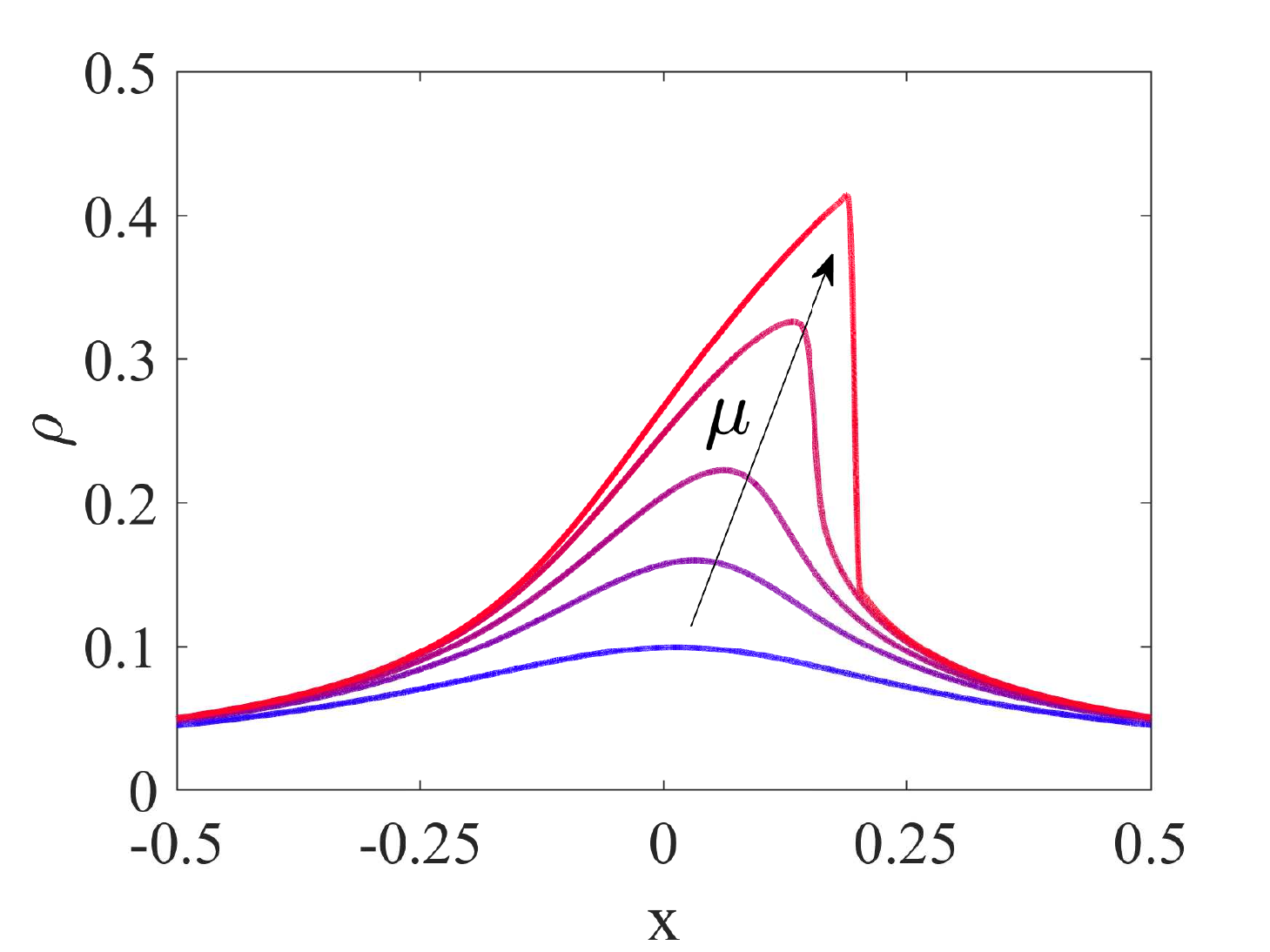}
		\caption{Profile amplitudes computed using~\eqref{eq:FCGLmodel} for additive forcing ($\Gamma_p=0$): (a) $\Gamma_a=0.001, 0.01, 0.05, 0.1$ at $\mu=-0.05$, and (b) $\mu=-1$, -0.6, -0.4, -0.2, -0.05 at $\Gamma_a=0.1$. Other parameters: $\beta=5$, $D=10^{-6}$, $\eta=2$.}
		\label{fig:profilesVsGaMu}
	\end{figure}

	While the above results have been obtained for additive forcing, similar ideas apply also to parametric and combined forcing. Moreover, qualitatively similar results are also obtained by varying other parameters, such as the forcing amplitude ($\Gamma_a$) and the distance from the Hopf bifurcation ($\mu$). Panels (a) and (b) of Fig.~\ref{fig:profilesVsGaMu} show how the shape becomes increasingly asymmetric when $\Gamma_a$ and $\mu$ are, respectively, increased. In both cases $\beta$ was kept constant.
	
	\section{Conclusion}\label{sec:discussion}
	
	In this work, we studied the shape of localized resonant oscillations that emerge in 1:1 periodically forced oscillatory systems with monotonic variations of natural frequencies in space. For simplicity, we confined ourselves in this study to a linear spatial dependence of the natural oscillation frequency, expressed in terms of the detuning parameter $\nu$. We focused on the role of bistability in shaping the asymmetry of localized profiles, using the nonlinear frequency correction, $\beta$, as a control parameter (see ~\eqref{eq:FCGLmodel}). The results indicate that bistability in a weakly coupled oscillatory system leads to an abrupt decline of the oscillation amplitude at a specific fixed location in space, related to the so-called Maxwell point, $\nu_M$, at which front solutions of the spatially homogeneous system are stationary. This location affects other properties of the spatial profile of resonant oscillations, such as the maximum amplitude, the width, and the profile symmetry,	as Fig.~\ref{fig:weakCouplingProfilesVsBeta} shows.
	
	We believe that a qualitative understanding of the resonant-profile asymmetry, as obtained in this study, can shed new light on the puzzling mechanisms that shape localized resonant oscillations in the cochlea~\cite{reichenbach2010ratchet}. Specifically, our results suggest that processes that lead to nonlinear frequency shifts ($\beta\neq0$), or affect the threshold of oscillations ($\mu$), can play primary roles in shaping the profile.
	The analysis reported here can be extended to include spatial variations in additional parameters, e.g. $\mu$, which is related	to quality factors of the oscillator~\cite{shera2002revised,bell2012resonance}, or the coupling between the outer hair cell and the basilar membrane~\cite{reichenbach2010ratchet,olson2012bekesy,Hudspeth2014,ammari2019fully,fallah2019nonlinearity}.
	
	More broadly, knowledge about the factors that shape localized oscillations can be used in various technological applications, by custom tailoring the spatial heterogeneity of the related media~\cite{moriarty2011faraday,urra2019localized}. Examples of such applications include mechanical resonators (NEMS and MEMS)~\cite{lifshitz2010nonlinear,jia2013parametrically,abrams2014nonlinear}, catalytic surface reactions~\cite{imbihl1995oscillatory} and plasmonic architectures~\cite{noskov2012oscillons}.
	An example of a more diverse application is plant communities subjected to seasonal forcing, where plant species constitute damped oscillators distributed inhomogeneously in trait space~\cite{Nathan2016j_ecology,tzuk2019interplay,tzuk2019period}. The properties of the profile of localized oscillations studied here can be related to community-level properties such as community composition (profile location), functional diversity (profile width), and resilience to environmental changes (profile asymmetry).

	\section*{Acknowledgments}
	This research was financially supported by the Ministry of Science and Technology of Israel (grant no. 3-14423) and the Israel Science Foundation under Grant No. 1053/17. Y.E. also acknowledges support by the Kreitman Fellowship.
	
	\providecommand{\noopsort}[1]{}\providecommand{\singleletter}[1]{#1}%

\end{document}